\documentclass[prd,aps,a4,twocolumn,superscriptaddress,preprintnumbers,nofootinbib]{revtex4-1}
\usepackage{pslatex}
\usepackage[pdftex]{graphicx}
\usepackage{psfrag}
\usepackage{epsfig}
\usepackage{color}
\usepackage{cancel}
\usepackage{slashed}
\usepackage{amssymb}
\usepackage{amsmath}
\usepackage{hyperref}
\usepackage{enumerate}
\usepackage{multirow}
\usepackage{ulem}

\allowdisplaybreaks
\begin{document}
\begin{flushright}
MI-TH-213
\end{flushright}
\title{Coherent elastic neutrino-nucleus scattering with the $\nu$BDX-DRIFT directional detector at next generation neutrino facilities}
\author{D. Aristizabal Sierra}%
\email{daristizabal@ulg.ac.be}%
\affiliation{Universidad T\'ecnica
  Federico Santa Mar\'{i}a - Departamento de F\'{i}sica\\
  Casilla 110-V, Avda. Espa\~na 1680, Valpara\'{i}so, Chile}%
\affiliation{IFPA, Dep. AGO, Universit\'e de Li\`ege, Bat B5, Sart
  Tilman B-4000 Li\`ege 1, Belgium}%
\author{Bhaskar Dutta}%
\email{dutta@physics.tamu.edu}%
\affiliation{Department of Physics and Astronomy, Mitchell Institute
  for Fundamental Physics and Astronomy, Texas A\&M University,
  College Station, TX 77843, USA}%
\author{Doojin Kim}%
\email{doojin.kim@tamu.edu} \affiliation{Department of Physics and
  Astronomy, Mitchell Institute for Fundamental Physics and Astronomy,
  Texas A\&M University, College Station, TX 77843, USA}%
\author{Daniel Snowden-Ifft}%
\email{ifft@oxy.edu}%
\affiliation{Occidental College, Dept. of Physics, 1600 Campus Road,
  Los Angeles, CA, United States}%
\author{Louis E. Strigari}%
\email{strigari@tamu.edu}%
\affiliation{Department of Physics and Astronomy, Mitchell Institute
  for Fundamental Physics and Astronomy, Texas A\&M University,
  College Station, TX 77843, USA}%
\begin{abstract}
  We discuss various aspects of a neutrino physics program that can
  be carried out with the neutrino Beam-Dump eXperiment DRIFT
  ($\nu$BDX-DRIFT) detector using neutrino beams produced in next generation neutrino facilities. $\nu$BDX-DRIFT is a directional low-pressure TPC
  detector suitable for measurements of coherent elastic
  neutrino-nucleus scattering (CE$\nu$NS) using a variety of gaseous
  target materials which include carbon disulfide, carbon tetrafluoride
  and tetraethyllead, among others. The neutrino physics program includes
  standard model (SM) measurements and beyond the standard model (BSM) physics
  searches. Focusing on the Long Baseline Neutrino Facility (LBNF) beamline at Fermilab, we first discuss basic features of the detector and estimate backgrounds, including beam-induced neutron backgrounds. We then quantify
  the CE$\nu$NS signal in the different target materials and study the
  sensitivity of $\nu$BDX-DRIFT to measurements of the weak mixing angle
  and neutron density distributions. We consider as well prospects
  for new physics searches, in particular sensitivities to effective
  neutrino non-standard interactions.
\end{abstract}

\maketitle
\section{Introduction}
\label{sec:intro}
Coherent elastic neutrino-nucleus scattering (CE$\nu$NS) is a process in which neutrinos scatter on a nucleus which acts as a single particle. Within the Standard Model (SM), CE$\nu$NS is fundamentally described by the neutral current interaction of neutrinos and quarks, and due to the nature of SM couplings it is approximately proportional to the neutron number squared~\cite{Freedman:1973yd}. Following years of experimental efforts, the COHERENT collaboration has established the first detection of CE$\nu$NS using a stopped-pion source with both a CsI{[Na]} scintillating crystal detector~\cite{Akimov:2017ade} and single-phase liquid argon target~\cite{Akimov:2020pdx}.

There are many proposed experimental ideas to follow up on the detection of CE$\nu$NS, using for example  reactor~\cite{Wong:2006nx,Billard:2016giu,Agnolet:2016zir,Ko:2016owz,Aguilar-Arevalo:2019jlr,Angloher:2019flc,Akimov:2019ogx,Fernandez-Moroni:2020yyl}, SNS~\cite{Akimov:2018ghi,Baxter:2019mcx}, and $^{51}$Cr sources~\cite{Bellenghi:2019vtc}. The COHERENT data and these future detections provide an exciting new method to study beyond the SM (BSM) physics through the neutrino sector, as well as provide a new probe of nuclear properties. 

Since the power of CE$\nu$NS as a new physics probe is just now being realized, it is important to identify new ways to exploit CE$\nu$NS in future experiments. In this paper, we propose a new idea to study CE$\nu$NS with the neutrino Beam-Dump eXperiment Directional Identification From Tracks ($\nu$BDX-DRIFT) detector using neutrino beams at next generation neutrino experiments. For concreteness we focus on the Long Baseline Neutrino Facility (LBNF) beamline at Fermilab \cite{Strait:2016mof}. As we show, this experimental setup is unique relative to on-going CE$\nu$NS experiments, for two primary reasons. First, the LBNF beam neutrinos are produced at a characteristic energy scale different than neutrinos from reactor or SNS sources. This provides an important new, third energy scale at which the CE$\nu$NS cross section can be studied. Second, our detector has directional sensitivity, which improves background discrimination and signal extraction. Previous studies have shown how directional sensitivity improves sensitivity for BSM searches~\cite{Abdullah:2020iiv}.  

This paper is organized as follows. In Section~\ref{sec:nuBDX-DRIFT-features} we discuss the basic features of the $\nu$BDX-DRIFT detector setup that we are considering. In Section~\ref{sec:general} we discuss the expected CE$\nu$NS signal at $\nu$BDX-DRIFT. In Section~\ref{sec:physics-analysis}A, we investigate the backgrounds at  $\nu$BDX-DRIFT and in Section~\ref{sec:physics-analysis}B, we show the aspects of SM and BSM physics that can be studied using $\nu$BDX-DRIFT. In Section~\ref{sec:conclusions} we present our conclusions.

\section{$\nu$BDX-DRIFT: Basic detector features}
\label{sec:nuBDX-DRIFT-features}
\begin{figure}
    \includegraphics[width=9.5cm]{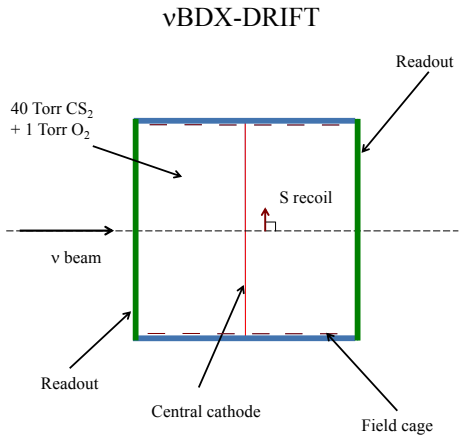}
    \caption{A sketch of the $\nu$BDX-DRIFT detector.
    \label{fig:vBDXDRIFTSketch}}
\end{figure}

As discussed in~\cite{Snowden-Ifft:2018bde}, a BDX-DRIFT detector, with its novel directional and background rejection capabilities, is ideally suited to search for elastic, coherent, low-energy, nuclear-recoils from light dark matter (DM). A sketch of a BDX-DRIFT detector is shown in Figure~\ref{fig:vBDXDRIFTSketch}. The readouts on either end couple to two back-to-back drift volumes filled with a nominal mixture of 40 Torr CS$_2$ and 1 Torr O$_2$ and placed into a neutrino beam, as shown. The use of the  electronegative gas CS$_2$ allows for the ionization to be transported through the gas with only thermal diffusion which largely preserves the shape of the track~\cite{SnowdenIfft:2013iy}. CS$_2$ releases the electron near the gain element allowing for normal electron avalanche to occur at the readout~\cite{SnowdenIfft:2013iy}. The addition of O$_2$ to the gas mixture allows for the distance between the recoil and the detector to be measured without a $t_0$ (time of creation of the ionization)~\cite{Battat:2016xxe,Snowden-Ifft:2014taa,Battat:2014van} eliminating, with side-vetoes, prodigious backgrounds from the edges of the fiducial volume. Because of the prevalence of S in the gas and the $Z^2$ dependence for coherent, elastic, low-energy scattering,~\cite{Snowden-Ifft:2018bde} the recoils would be predominantly S nuclei.  With a theshold of 20 keV the S recoils would be scattered within one degree of perpendicular to the beam line due to extremely low-momentum transfer, scattering kinematics. The signature of these interactions, therefore, would be a population of events with ionization parallel to the detector readout planes.

Here we consider deploying a BDX-DRIFT detector in a neutrino beam of a next generation neutrino facility, which for definitiveness we take to be the LBNF beamline at Fermilab. As discussed below CE$\nu$NS will produce low-energy nuclear recoils in the fiducial volume of a $\nu$BDX-DRIFT detector. To optimize the detector for CE$\nu$NS detection various gas mixtures and pressures are considered.

\section{CE$\nu$NS in $\nu$BDX-DRIFT}
\label{sec:general}
\begin{figure*}
    \centering
    \includegraphics[scale=0.4]{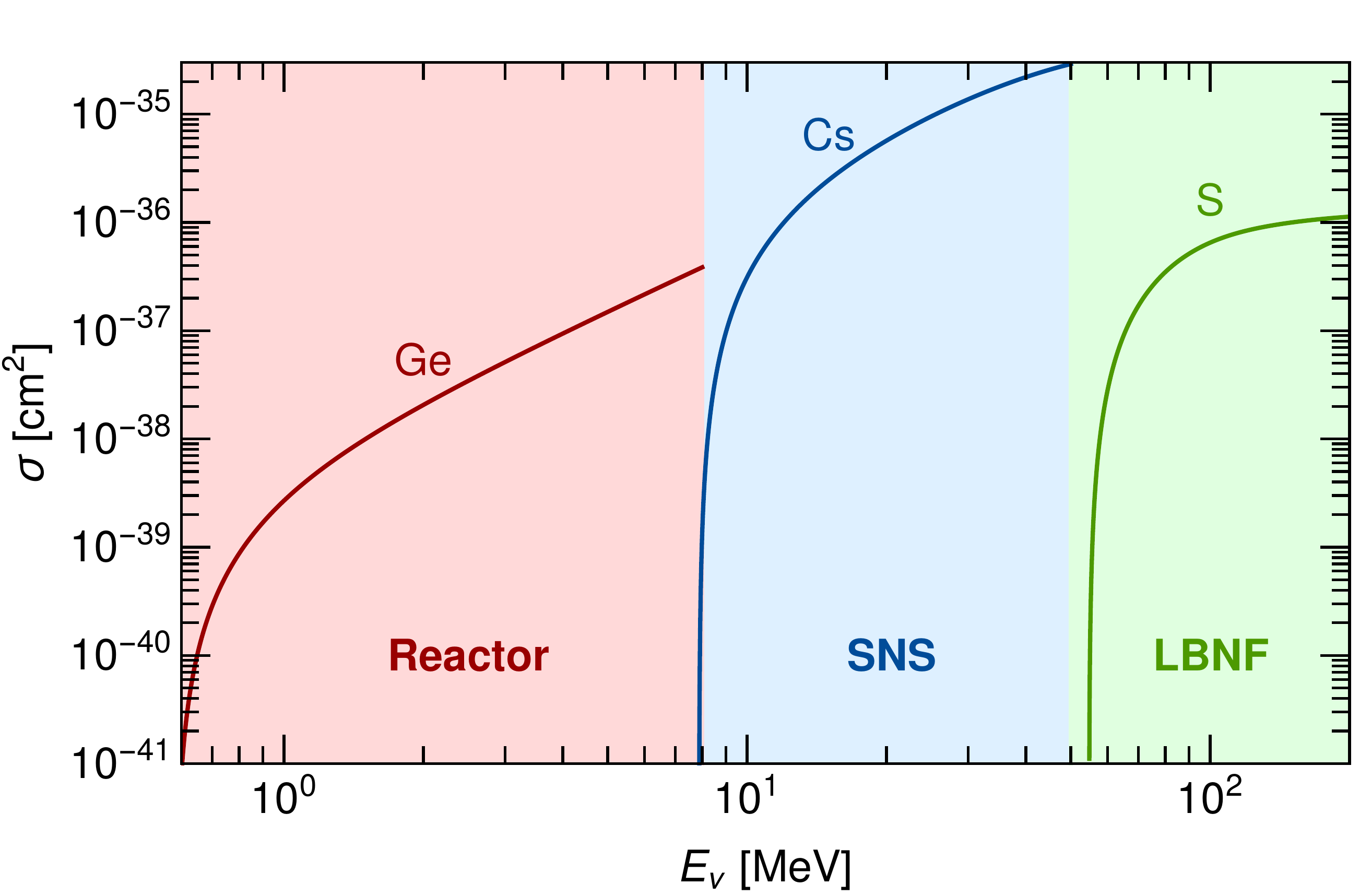}
    \caption{CE$\nu$NS total cross section as a function of incoming 
    neutrino energy for reactor neutrinos, spallation neutron source (SNS)
    neutrinos and the LBNF beamline. Cross sections are calculated for
    representative nuclides of the technologies used in each case:
    germanium (reactor), cesium (SNS) and sulfur (LBNF). This graph shows the
    different energy domains at which a significant CE$\nu$NS signal can be induced.}
    \label{fig:cevns_energy_domains}
\end{figure*}

When the neutrino-nucleus exchanged momentum is small enough
($q\lesssim 200\,$MeV) the individual nucleon amplitudes sum up
coherently, resulting in a coherent enhancement of the
neutrino-nucleus cross section \cite{Freedman:1973yd}. So rather than
scattering off nucleons the neutrino scatters off the entire nucleus.
This constraint on $q$ translates into an upper limit on the neutrino
energy $E_\nu\lesssim 100\,$MeV, which in turn ``selects'' the
neutrino sources capable of inducing CE$\nu$NS. At the laboratory
level, reactor neutrinos with $E_\nu\lesssim 9\,$MeV dominate the low
energy window, while stopped-pion sources with $E_\nu<m_\mu/2$ the
intermediate energy window. Fig. \ref{fig:cevns_energy_domains} shows the
different energy domains at which CE$\nu$NS can be induced. At the 
astrophysical level CE$\nu$NS can be instead induced by solar, supernova
and atmospheric neutrinos in the low, intermediate and ``high'' energy
windows, respectively.

Using laboratory-based sources, CE$\nu$NS has been measured by the
COHERENT collaboration with CsI[Na] and LAr detectors
\cite{Akimov:2017ade,Akimov:2020czh}. And measurements using reactor
neutrino sources are expected in the near-future
\cite{Agnolet:2016zir,Aguilar-Arevalo:2019jlr,Strauss:2017cuu}. The
high-energy window however has been rarely discussed and experiments
covering that window have been so far not considered. One of the
reasons is probably related with the conditions that should be
minimally satisfied for an experiment to cover that energy range: (i)
The low-energy tail of the neutrino spectrum should provide a
sufficiently large neutrino flux, (ii) the detector should be
sensitive to small energy depositions and (iii) backgrounds need to be sufficiently small to observe the signal. The LBNF beamline combined with
the $\nu$BDX-DRIFT detector satisfy these three criteria, as we will now
demonstrate.

Accounting for the neutron and proton distributions independently,
i.e. assuming that their root-mean-square (rms) radii are different
$\langle r^2_n\rangle\neq \langle r^2_p\rangle$, the SM CE$\nu$NS
differential cross section reads
\cite{Freedman:1973yd,Freedman:1977xn}
\begin{equation}
  \label{eq:x-sec-cevns}
  \frac{d\sigma}{dE_r}=\frac{m_NG_F^2}{2\pi}
  \left(2-\frac{E_r m_N}{E_\nu^2}\right)Q^2_W\ ,
\end{equation}
where the coherent weak charge quantifies the $Z$-nucleus vector coupling, namely
\begin{equation}
    \label{eq:coherent-weak-charge}
    Q_W^2=\left[Ng_V^n F_N(q) + Zg_V^p F_Z(q)\right]^2\ .
\end{equation}
The proton and neutron charges are determined by the up and down quark
weak charges and read $g_V^n=-1/2$ and $g_V^p=1/2-2\sin^2\theta_W$. In
the Born approximation the nuclear form factors are obtained from the
Fourier transform of the neutron and proton density distributions. The
properties of these distributions are captured by different parametrizations,
which define different form factors. For all our calculations we use the one
provided by the Helm model \cite{Helm:1956zz}, apart from Section
\ref{sec:neutron-density-distributions} where we will as well consider those
given by the symmetrized Fermi distribution function and the Klein-Nystrand
approach \cite{0305-4470-30-18-026,Klein:1999qj} (see that Section for details).
Note that the dependence that the signal has on the form factor choice is a
source for the signal uncertainty.

In almost all analyses $\langle r^2_n\rangle=\langle r^2_p\rangle$, and
so the form factor factorizes. That approximation is good enough unless one is
concerned about percent effects
\cite{AristizabalSierra:2019zmy,Hoferichter:2020osn},
$\langle r^2_p\rangle$ values for $Z$ up to 96 are known at the
part per thousand level through elastic electron-nucleus scattering
\cite{Angeli:2013epw}. In that limit one can readily see that the
differential cross section is enhanced by the number of neutrons
($N^2$) of the target material involved, a manifestation of the
coherent sum of the individual nucleon amplitudes. In what follows all
our analyses will be done in that limit, the exception being Sec. \ref{sec:neutron-density-distributions}.

\begin{figure*}
  \centering
  \includegraphics[scale=0.33]{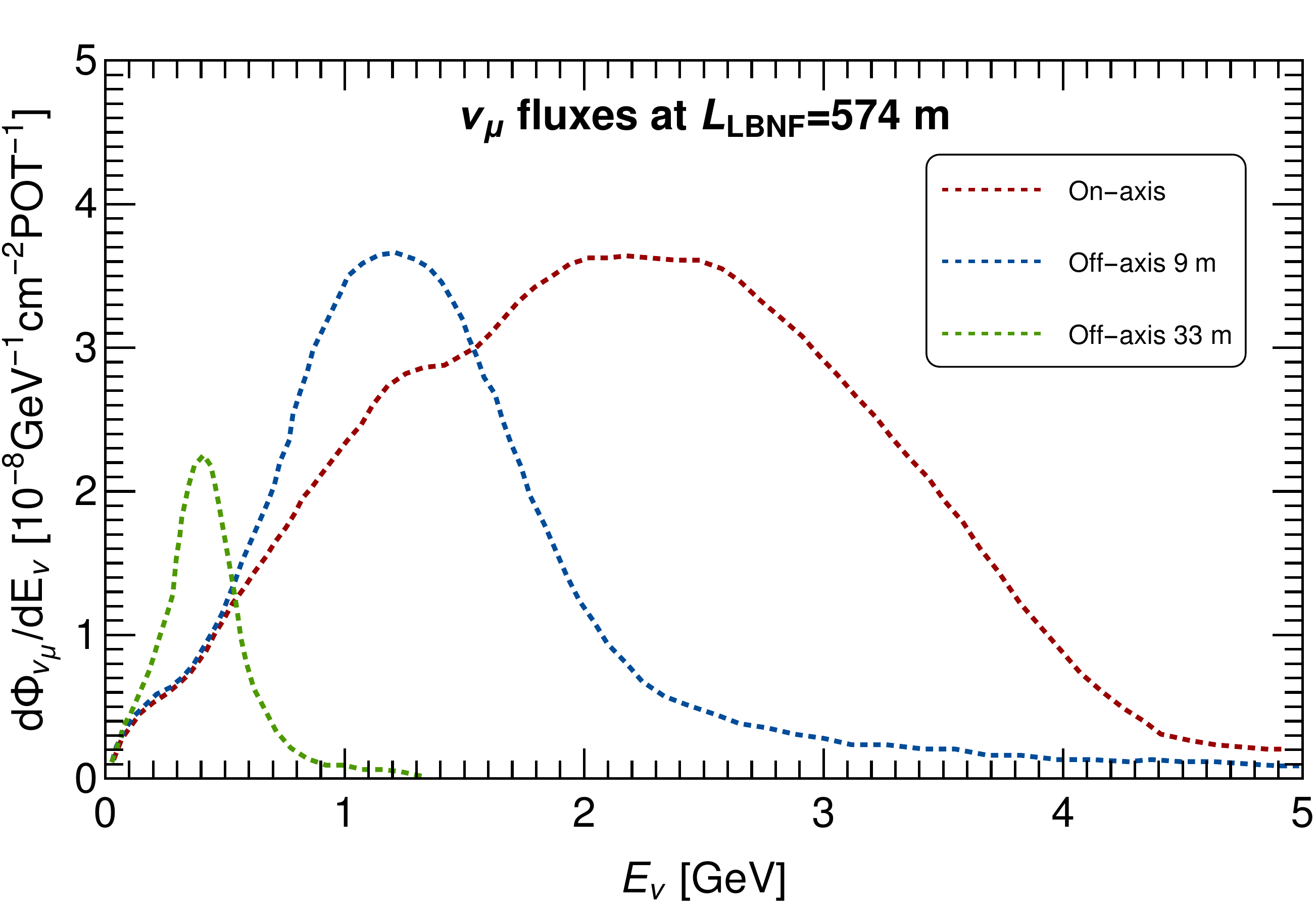}
  \includegraphics[scale=0.336]{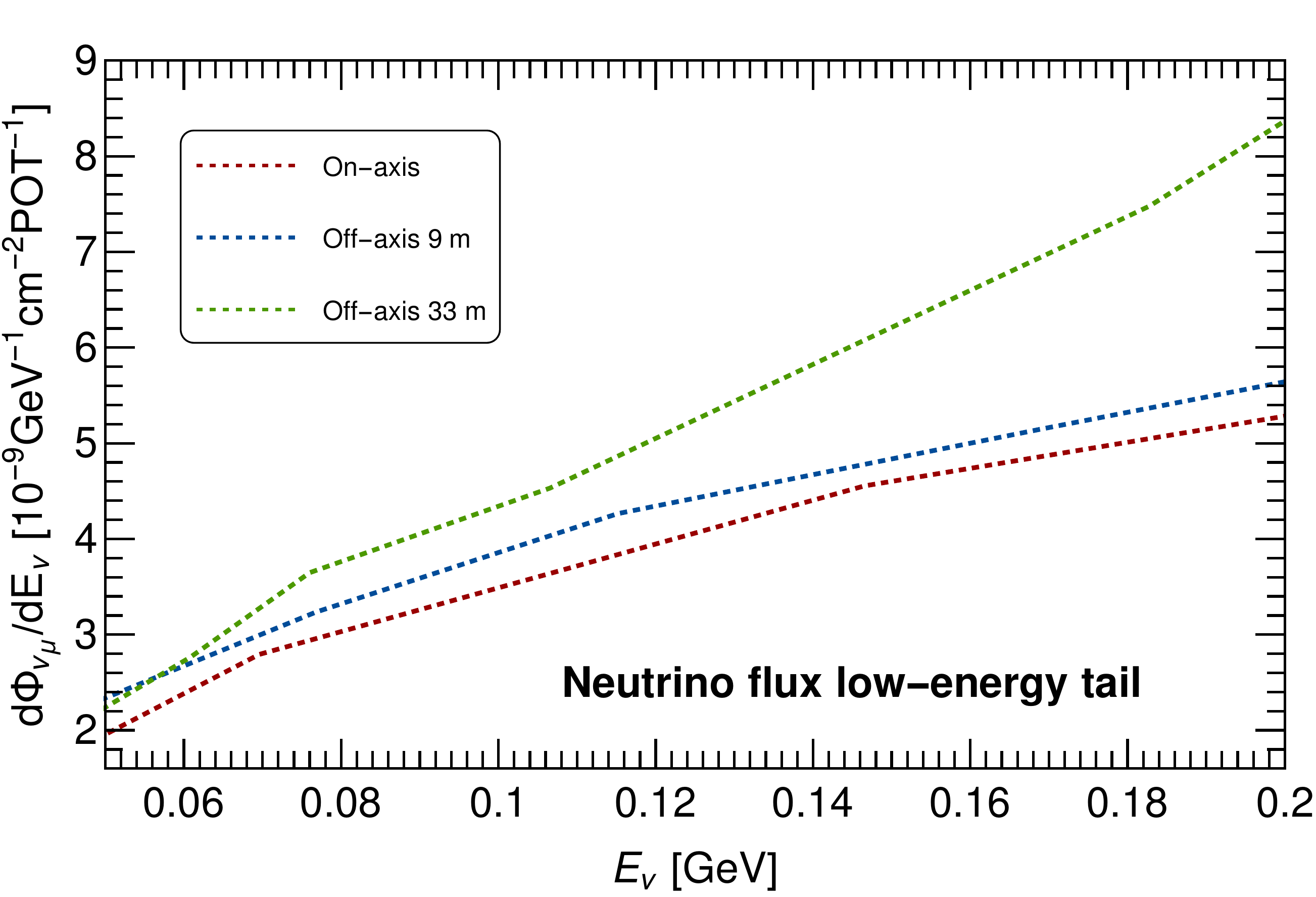}
  \caption{\textbf{Left graph}: Neutrino flux at the $\nu$BDX-DRIFT location ($L_\text{LBNF}=574\,$m) for three
    positions: on-axis and off-axis 9 m and 33 m ($0.5^\circ$ and
    $2.0^\circ$ off-axis)~\cite{Abi:2020evt}.
    \textbf{Right graph}: Low-energy tail of the neutrino spectra,
    relevant for CE$\nu$NS, for the three positions considered in the left graph.
    See discussion in Section \ref{sec:general} for details.}
  \label{fig:fluxes}
\end{figure*}

The differential event rate (events/year/keV) follows from a
convolution of the CE$\nu$NS differential cross section and the
neutrino spectral function properly normalized
\begin{equation}
  \label{eq:diff-event-rate}
  \frac{dR}{dE_r}=V_\text{det}\,\rho(P)
  \frac{N_A}{m_\text{molar}}\int_{E_\nu^\text{min}}^{E_\nu^\text{max}}
  \,\frac{d\sigma}{dE_r}\,\frac{d\Phi}{dE_\nu}dE_\nu\ .
\end{equation}
Here $E_\nu^\text{min}=\sqrt{m_N E_r/2}$. The first two factors define
the detector mass $m_\text{det}=V_\text{det}\,\rho(P)$, where
$\rho(P)$ corresponds to the target material density which depends on
detector pressure at fixed room temperature, $T=293\,$K. Assuming an
ideal gas it reads,
\begin{equation}
  \label{eq:density}
  \rho=5.5\times10^{-5}\times
  \left(\frac{m_\text{molar}}{\text{g}/\text{mol}}\right)
  \left(\frac{P}{\text{Torr}}\right)
  \frac{\text{kg}}{\text{m}^3}\ .
\end{equation}
Pressure and recoil energy threshold are related and their dependence
varies with target material. For the isotopes considered here,
assuming CS$_2$ to be the dominant gas, we have:
\begin{equation}
    \label{Eq:thresholds-pressure-Er}
    E_r^\text{th}(\text{Nuc}_i)=f_i\,\left(\frac{P}{40~\text{Torr}}\right)
    \,\text{keV}\ ,
\end{equation}
with $f_i=\{2,7.5,13,20,69\}$ for $\text{Nuc}_i=\{\text{H},\text{C},\text{F},\text{S},\text{Pb}\}$~\cite{Battat:2016xxe,Burgos:2007zz}.
For the neutrino spectrum (and normalization) we use the DUNE near
detector flux prediction for three different positions (on-axis and
off-axis 9 m ($0.5^\circ$ off-axis) and 33 m ($2.0^\circ$ off-axis))~\cite{Abi:2020evt}. Fig. \ref{fig:fluxes} shows the corresponding
fluxes (left graph) along with the low energy region relevant for
CE$\nu$NS (right graph).

With these results we are now in a position to calculate the CE$\nu$NS
event yield for potential different target materials (compounds):
carbon disulfide, carbon tetrafluoride and tetraethyllead as a function of pressure (threshold). We start with carbon disulfide and assume the following
detector configuration/operation values: $V_\text{det}=10\,\text{m}^3$
and seven-year data taking. Results for smaller/larger detector volumes
as well as for smaller/larger operation times follow from an overall
scaling of the results presented here, provided the assumption of a
pointlike detector is kept.

\begin{figure*}
  \centering
  \includegraphics[scale=0.33]{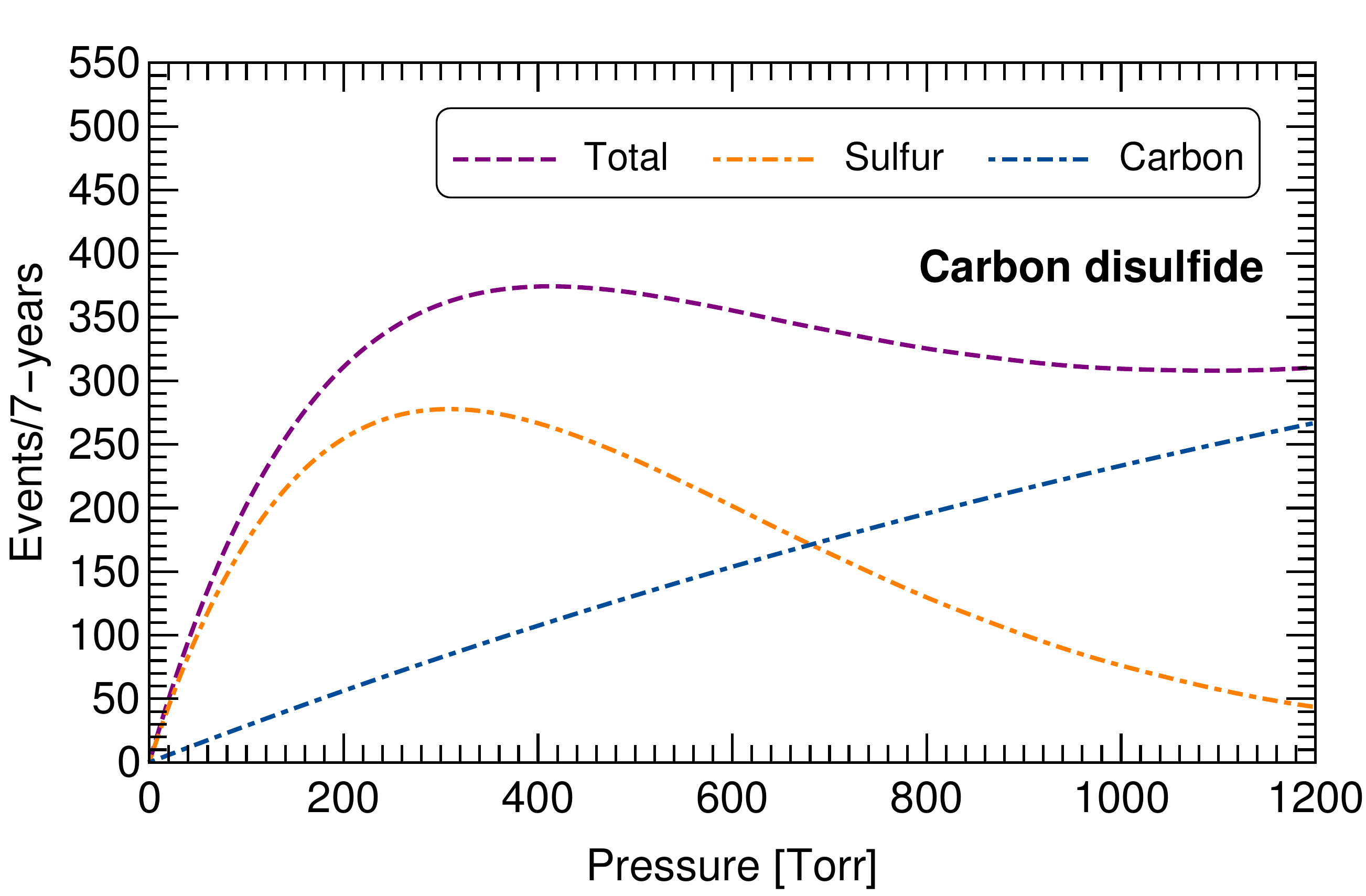}
  \includegraphics[scale=0.33]{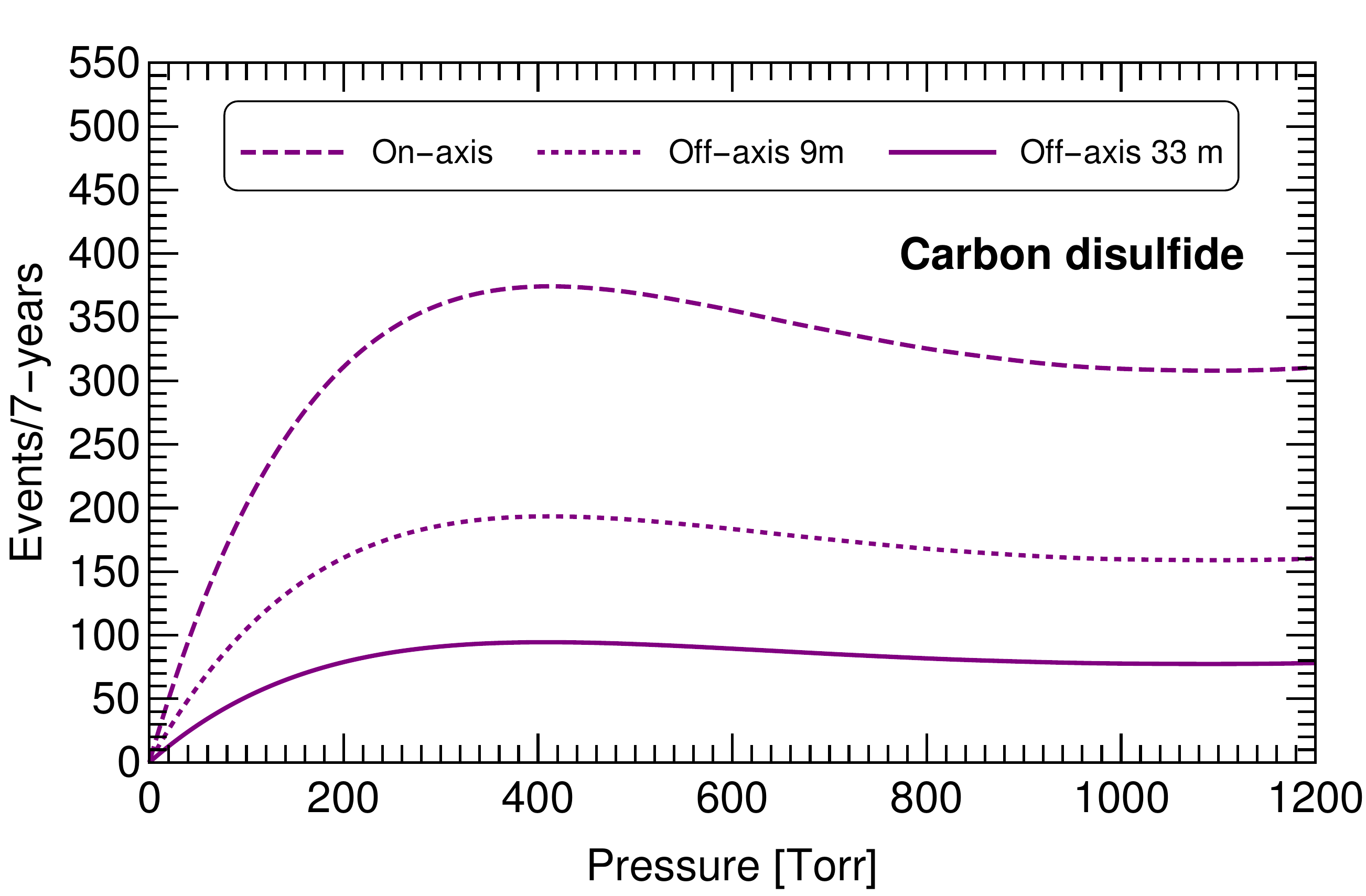}
  \caption{\textbf{Left graph}: CE$\nu$NS event yield for carbon
    dioxide in terms of pressure assuming a ten-cubic meter detector
    volume operating at room temperature during a 7-years data taking period. \textbf{Right graph}:
    CE$\nu$NS event yield for different detector position
    configurations: on-axis and off-axis $9\,$m and $33\,$m
    ($0.5^\circ$ and $2.0^\circ$ off-axis).}
  \label{fig:carbon-dioxide}
\end{figure*}

Left graph in Fig. \ref{fig:carbon-dioxide} shows the CE$\nu$NS event
rate for CS$_2$, carbon and sulfur independently displayed. The result
is obtained by assuming the on-axis neutrino flux configuration. One
can see that up to 700 Torr the event rate is dominated by the sulfur
contribution, point at which carbon overtakes the event rate with a
somewhat degraded contribution. The individual behavior of each
contribution can be readily understood as follows. At low recoil
energies the event rate is rather flat but pressure is low, thus the
suppression of both contributions in that region is due to low pressure. As
pressure increases, $m_\text{det}$ increases, as do the carbon and sulfur event rates. There is a pressure, however, for which the
processes start losing coherence and so the event rates start
decreasing accordingly (variations in pressure translate into
variations in recoil energy threshold according to Eq.
(\ref{Eq:thresholds-pressure-Er})). For sulfur it happens at lower pressures than
for carbon, as expected given that sulfur is a heavier nucleus. For
CS$_2$ then it is clear that the optimal pressure is set at about 400
Torr (exactly at 411 Torr), a value that corresponds to
$E_r^\text{th}\simeq 77.1\,$keV for carbon and to
$E_r^\text{th}=205.5\,$keV for sulfur, according to Eq. (\ref{Eq:thresholds-pressure-Er}).  In summary, at the optimum pressure and corresponding threshold, for CS$_2$ the number of CE$\nu$NS events for a 7-year 10 cubic-meter exposure is 367.

Although rather energetic, it is clear that the LBNF beamline can induce CE$\nu$NS and that the process can be measured, provided the
detector is sensitive to low recoil energies. The details of how
CE$\nu$NS proceeds are as follows. The low-energy tail of the neutrino
spectrum (on-axis) extends down to energies of order 50 MeV or so, as
can be seen in the right graph in Fig.~\ref{fig:fluxes}. From that
energy and up to those where coherence is lost, the neutrino flux will
induce a sizable number of CE$\nu$NS events. Taking the recoil energy
at which $F^2(E_r)$ decreases from $1$ to $0.1$ as the energy at which
coherence is lost (above those energies the nuclear form factor
decreases rapidly and enters a dip, regardless of the nuclei), $E_\nu$
can be determined with the aid of $E_\nu=\sqrt{2m_NE_r}$. Since for
sulfur (carbon) we found $E_r^\text{S}\simeq 370\,$keV
($E_r^\text{C}\simeq 1800\,$keV), we get
$E_\nu^\text{S}\simeq 150\,$MeV ($E_\nu^\text{C}\simeq 200\,$MeV).
Numerically we have checked that the event yield changes only in
one part per thousand when increasing $E_r^\text{max}$ to
values for which $E_\nu>200\,$MeV.

The number of muon neutrinos per year per $\text{cm}^2$ delivered by
the LBNF beamline in the on-axis configuration and the full energy range, $[0.5,5\times 10^3]\,$ MeV, is $1.1\times 10^{14}$. For the
energy range that matters for CE$\nu$NS this number is instead $10^{12}$. There are about two orders of magnitude less neutrinos for
CE$\nu$NS than e.g. for elastic neutrino-electron scattering. However,
the flux depletion is somewhat compensated by the $N^2$ enhancement of
the CE$\nu$NS cross sections, which for sulfur (carbon) amounts to
$256$ ($36$). Thus, although fewer neutrinos are available for
CE$\nu$NS, the large size of the corresponding cross section leads to a
sizable number of events even for neutrino energies far above those of
spallation neutron source neutrinos.

\begin{figure*}
  \centering
  \includegraphics[scale=0.34]{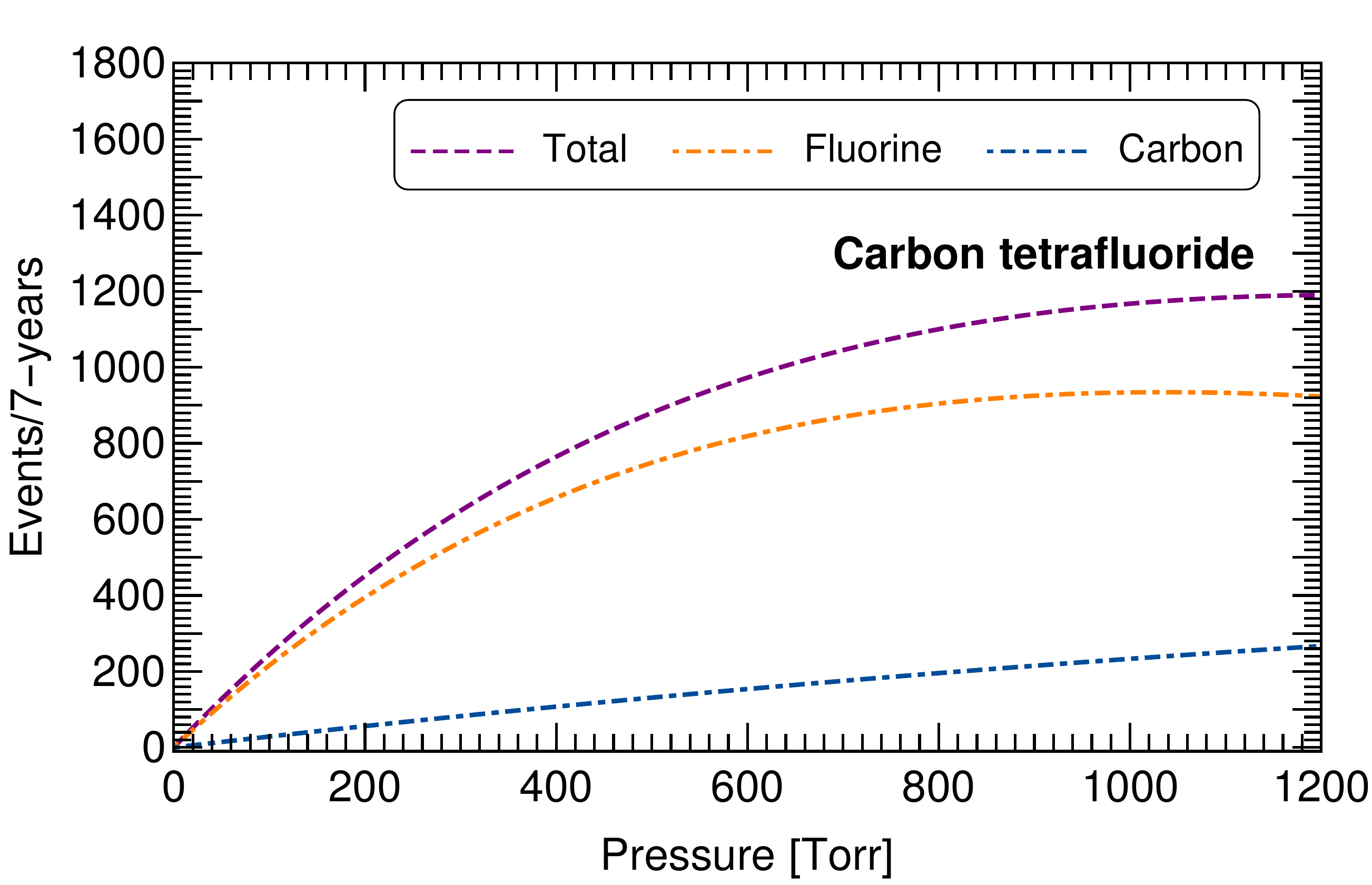}
  \includegraphics[scale=0.325]{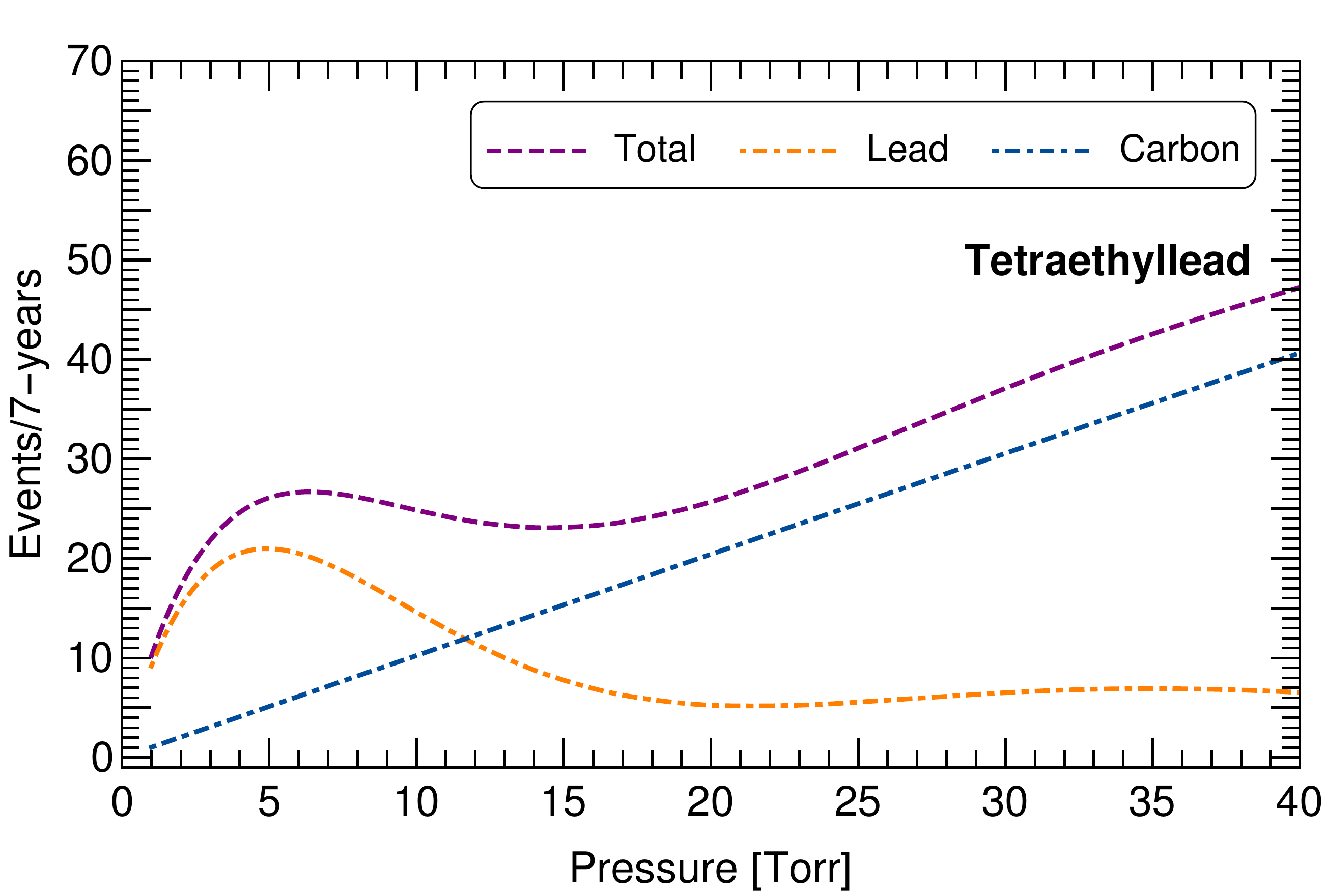}
  \caption{\textbf{Left graph}: CE$\nu$NS event rates as a function of
    pressure for carbon tetrafluorine ($\text{CF}_4$) assuming a ten-cubic
    meter detector volume, operation at room temperature and on-axis
    neutrino flux configuration. The calculation assumes the bulk of the
    gas is filled with $\text{CF}_4$. \textbf{Right graph}: Same as left
    graph but for tetraethyllead ($\text{C}_8\text{H}_{20}\text{Pb}$).
    Hydrogen contributes to the signal at the per mille level and so its
    contribution is not displayed. In contrast to $\text{CF}_4$, in this
    case a concentration of 2.3:1 of carbon disulfide and tetraethyllead
    has been assumed.}
  \label{fig:other-materials-cevns}
\end{figure*}
From Fig.~\ref{fig:fluxes}, and as expected, it is clear that the
number of neutrinos decreases as one moves off the axis. For the
configurations shown there we calculate:
and $n_\nu(33\text{m})=9.3\times
10^{12}\,\nu_\mu/\text{year}/\text{cm}^2$ (integrating over the
full neutrino energy range, $[5\times 10^{-2},5]\,$GeV).
So the CE$\nu$NS event rates for these off-axis configurations are
depleted, although as a function of pressure they keep the same
behavior, as can be seen in the right graph in
Fig. \ref{fig:carbon-dioxide}. Note that off-axis configurations, in
particular that at $33\,\text{m}$, can potentially be ideal for light DM
searches since they lead to a suppression of neutrino (or neutrino-related) backgrounds \cite{nubdx-drift-DM}.

$\nu$BDX-DRIFT is suitable for other target materials as well, so we
have investigated the behavior of their event rates. The left graph in
Fig.~\ref{fig:other-materials-cevns} shows the result for carbon
tetrafluoride (CF$_4$), while the right graph for tetraethyllead ($\text{C}_8\text{H}_{20}\text{Pb}$). For the results in the left 
graph we have assumed the bulk of the gas is filled with $\text{CF}_4$,
i.e. 100\% of the fiducial volume is filled with $\text{CF}_4$.
Note that this a rather good approximation given that $\text{CS}_2$
and $\text{CF}_4$ have about the same number of electrons per molecule.
For the results in the right graph we have instead taken a CS$_2$:$\text{C}_8\text{H}_{20}\text{Pb}$ concentration of 2.3:1.
As we will discuss in Section \ref{sec:neutron-density-distributions},
these compounds are particularly useful for measurements of the
root-mean-square (rms) radius of the neutron distributions of carbon,
fluorine and lead.

From these results one can see that for carbon tetrafluoride the
signal is dominated by fluorine, with subdominant contributions
from carbon. Fluorine being a slightly heavier nuclei has intrinsically
a larger cross section, with an enhancement factor of order
$(N_\text{F}/N_\text{C})^2=100/36\simeq 2.8$. In addition the
carbon-to-fluorine ratio of the compound implies an extra factor 4 for
the fluorine contribution. One can see as well that up to 1200 Torr
the signal increases. For analyses in $\text{CF}_4$ we take the CE$\nu$NS
signal at $400\,$Torr, for which we get 808 events/7-years.

In terms of pressure, tetraethyllead behaves rather
differently. The signal is dominated by lead up to 12 Torr or so.
At that point the carbon contribution kicks in and dominates the signal,
particularly at high pressure. Hydrogen contributes to the signal at
the per mille level, despite being enhanced by a factor 20
from the molecular composition. This is expected, in contrast
to the carbon and lead cross sections the hydrogen contribution is not
enhanced. The behavior of the lead and carbon contributions can be
readily understood. Relative to lead the carbon coherence enhancement
factor is small $(N_\text{C}/N_\text{Pb})^2\simeq 2.3\times 10^{-3}$.
However, lead loses coherence at rather low pressures and so the difference
is mitigated. One can see that for $P<12\,$Torr carbon contributes
at the percent level.

The pressure at which the lead signal peaks is relevant if one is
interested in lead related quantities. That pressure corresponds to 
$6.4\,$Torr, for which carbon contributes about $25\%$ of the total
signal. At that pressure the signal amounts to 26 events/7-years, with the contribution from lead (carbon) equal to 19.2 events/7-years (6.7 events/7-years). Thus for such measurements one will need as well to distinguish
recoils in lead from those in carbon, something that seems viable given
that the range of C for a given ionization should be much larger than for
Pb.
\section{$\nu$BDX-DRIFT physics potential beyond CE$\nu$NS measurements}
\label{sec:physics-analysis}
After discussing CE$\nu$NS measurements with the $\nu$BDX-DRIFT
detector, we now proceed with a discussion of possible problematic
backgrounds as well as studies that
can be carried out with the detector. For the latter we split the
discussion in measurements of SM quantities and BSM searches. We would like to stress that although BSM searches at $\nu$BDX-DRIFT include those for light DM, here we limit our discussion to the case of new interactions in the neutrino sector that can potentially affect the CE$\nu$NS event spectrum. The discussion of light DM will be presented elsewhere \cite{nubdx-drift-DM}.
\subsection{Estimation of backgrounds at $\nu$BDX-DRIFT}
\label{sec:backgrounds-nuBDX-DRIFT}
DRIFT detectors have been shown to be insensitive to all types of
ionizing radiation except nuclear recoils after analysis cuts have
been applied with minimal loss of
sensitivity~\cite{Battat:2016xxe}. The most recent results from the
Boulby Mine show no nuclear recoil events in the fiducial volume in 55
days of running~\cite{Battat:2016xxe}.  These results have been
extended now to 150 days of running~\cite{SnowdenIfft:2021b}.
Furthermore DRIFT detectors have been run on the surface and only been
found to be sensitive to cosmic ray neutrons~\cite{SnowdenIfft:2021a}. The DUNE near detector site is at a depth of 60 m and any possibility of nuclear recoils induced from
cosmic rays at this shallower depth than the Boulby Mine would be
vetoed by timing cuts.  The several second cycle time of the LBNF beam
is ideally suited to the slow drift speed of a DRIFT detector.  Thus
beam-unrelated backgrounds will not be a limitation at LBNF.

Beam-related backgrounds are possibly concerning.  In this section we address the most worrisome beam-related background, neutrino-induced neutrons (NINs). The neutrino beam interacts not only with the target material but with the vessel walls as well. In that process, some neutrinos can interact with the nucleons of the vacuum vessel to produce neutrons, which could enter the active detector volume and produce a background of low-energy nuclear-recoils. Depending on the neutrino beam energy distribution, and the vacuum vessel material, different processes are to be considered. For an iron vessel (mostly $^{56}$Fe) and $E_\nu\lesssim 0.1\,$GeV, the incoming neutrino can strip off a neutron from $^{56}$Fe, thus inducing the stripping reaction $^{56}\text{Fe} + \nu_\mu\to n + ^{55}\text{Fe} + \nu_\mu$. The total cross section for this processes ranges from $10^{-42}\,\text{cm}^2$ to $10^{-41}\,\text{cm}^2$, and dominates NIN production in that neutrino energy regime \cite{Kolbe:2000np}.

For neutrino energies above $\sim 0.1\,$GeV other processes can dominate. The on-axis LBNF spectrum peaks within 2-3 GeV and extends up to energies of order 5 GeV (see Fig. \ref{fig:fluxes}). Thus, although LBNF neutrinos trigger iron stripping reactions, their rate is small compared to neutrino processes which open up as soon as $E_\nu\gtrsim 0.1\,$GeV, namely: elastic scattering (E); quasielastic scattering (QE); resonant single pion production (RES); deep inelastic scattering (DIS).\footnote{Coherent pion production, multipion production and kaon production open up as well at these energies, however their total cross sections are smaller \cite{Formaggio:2013kya}.} Of course, not all these processes produce final state neutrons, only E and RES do. For initial-state neutrinos, RES processes are \cite{Formaggio:2013kya}
\begin{alignat}{2}
\label{eq:CC-RES}
\text{CC:}\quad & \nu_\mu + p\to \mu^- + p + \pi^+\ , 
&\quad \nu_\mu + n\to \mu^- + p + \pi^0 \ ,
\nonumber\\
&\nu_\mu + n\to \mu^- + n + \pi^+\ ,
\\
\label{eq:NC-RES}
\text{NC:}\quad& \nu_\mu + p\to \nu_\mu + p + \pi^0\ ,
&\quad \nu_\mu + p\to \nu_\mu + n + \pi^+\ ,
\nonumber\\
& \nu_\mu + n\to \nu_\mu + n + \pi^0\ ,
& \nu_\mu + n\to \nu_\mu + p + \pi^-\ .
\end{alignat}
Thus, only three out of seven involve final-state neutrons which could give recoils mimicking the signal. As can be seen in Eq. (\ref{eq:CC-RES}) RES CC processes produce as well charged products which would likely be picked up in the fiducial volume and so vetoed, but are included here for a generous estimate of the backgrounds.  The protons produced by the other processes could produce recoils but are charged and so could be similarly vetoed.  Pions are either charged and so can be vetoed or uncharged and decaying so quickly to photons that they cannot produce recoils. So we then estimate the $\nu \to n$ total cross section according to
\begin{equation}
    \label{eq:nu-neutron-xsec}
    \sigma_\text{NIN}=\sigma_\text{E} + \frac{3}{7}\sigma_\text{RES}\ ,
\end{equation}
where we assume that the seven RES processes contribute equally to the RES total cross section. Fixing $E_\nu=3\,$GeV and
using the SM prediction for the total neutrino cross section at these energies \cite{Formaggio:2013kya} one then gets $\sigma_\text{NIN}=6.2\times 10^{-39}\,\text{cm}^2$.

With the relevant cross section estimated we can now calculate the expected number of NIN events. Assuming the full $\nu$BDX-DRIFT detector will be made of $N_\text{modules}$  $\nu$BDX-DRIFT modules each having $1\,\text{m}^3$ fiducial volumes surrounded by vacuum vessels $150\,\text{cm}$ on a side we then write the number of NIN per cycle as follows
\begin{align}
    \label{eq:number-of-NIN-per-cycle}
    \frac{N_\text{NIN}}{\text{cycle}}=&
    3.0\cdot 10^{-4}\left(\frac{\mathcal{F}}{3}\right)
    \left(\frac{n_\text{Fe}}{2.4\cdot 10^{24}/\text{cm}^3}\right)
    \left(\frac{n_\nu}{772640/\text{cm}^2}
    \right)
    \nonumber\\
    &\left(\frac{A}{22500\,\text{cm}^2}\right)
    \left(\frac{t}{1\,\text{cm}}\right)
    \left(\frac{\sigma_\text{NIN}}{6.2\cdot 10^{-39}\,\text{cm}^2}\right)
    \,N_\text{modules}\ .
\end{align}
Here $\mathcal{F}$ refers to the number of faces, $n_\text{Fe}$ to the iron neutron density, $n_\nu$ to the number of neutrinos per cycle, $A$ to the area of each face, and $t$ to the vessel wall thickness. We assume only 3 detector faces (front and half of the four lateral faces) are relevant because of forward scattering of the neutrons, while for $n_\nu$ we take the on-axis neutrino flux in Fig. \ref{fig:fluxes} rescaled by $n_\text{POT}/\text{cycle}=7.5\times 10^{13}$ \cite{Strait:2016mof}. Taking
1.0 second as a representative LBNF cycle time (LBNF extractions oscillate in the range 0.7-1.2 s \cite{Strait:2016mof}), a 10 cubic-meter detector and a data-taking period of 7 years, one gets $N_\text{NIN}/\text{7-\text{years}}\simeq 6.56\times 10^5$.

Given the LBNF beamline energy spectrum and the final-state particles in the processes of interest (E and RES), NINs are order GeV. The detection probability $\mathcal{P}$ for those GeV neutrons at $\nu$BDX-DRIFT operating with 100\% of the fiducial volume filled with CS$_2$ at $400\,\text{Torr}$ has been determined by a \texttt{GEANT4}~\cite{Agostinelli:2002hh} simulation benchmarked to neutron-induce nuclear-recoil data~\cite{Battat:2016xxe}. The result is $\mathcal{P}=2.5\times 10^{-5}$. With this number we then estimate the number of effective NIN events over the relevant time period and for 10 modules to be
\begin{align}
    \label{eq:NIN-data-taking-time}
    \frac{N_\text{NIN}^\text{eff}}{\text{7-years}}=&\mathcal{P}\times \left(\frac{N_\text{NIN}}{\text{7-\text{years}}}\right)
                                            \nonumber\\
                                            =&16.0\ .
\end{align}

From the CS$_2$ calculation represented in the left graph of Fig. \ref{fig:carbon-dioxide} we expect 367 CE$\nu$NS events above threshold for the same exposure.
This means that the signal-to-background (NIN) ratio is about 23, a number comparable to what the COHERENT collaboration found for the same type of events (47) \cite{Akimov:2017ade}. Following this analysis, our conclusion is that the NIN background contamination of the CE$\nu$NS signal is small for all possible realistic detector configurations.

NINs produced in the surrounding environment are less concerning as they can be shielded against either passively or actively, e.g. \cite{Westerdale:2016}.

\subsection{SM and BSM studies with the $\nu$BDX-DRIFT detector}
\label{sec:dir-signals}
Measurements of the CE$\nu$NS event spectrum can be used to extract
information on the weak mixing angle as well as on the rms radii of
neutron distributions. Using COHERENT CsI[Na] and LAr data this
approach has been used for $\sin^2\theta_W$
\cite{Kosmas:2017tsq,Miranda:2020tif}. It has been used as well in
forecasts of near-future reactor-based CE$\nu$NS data
\cite{Canas:2016vxp,Canas:2018rng}. These analyses provide relevant
information for this SM parameter at renormalization scales of order
$\langle q\rangle=\mu\simeq 10^{-2}\,$GeV and
$\mu\simeq 10^{-3}\,$GeV, respectively. An analysis complementary to
CE$\nu$NS-related experiments has been as well discussed using elastic
neutrino-electron scattering with the DUNE near detector
\cite{deGouvea:2019wav}. This measurement will provide information at
$\mu\simeq 6\times 10^{-2}\,$GeV, with
higher precision that what has been so far obtained by COHERENT and
comparable to what will be obtained with e.g. MINER and CONNIE.

Measurements of the rms radii of neutron distributions can be as well
performed through the observation of the CE$\nu$NS process. From
Eq. (\ref{eq:x-sec-cevns}) one can see that information on the
CE$\nu$NS event spectrum can be translated into limits on
$r_\text{rms}^n=\sqrt{\langle r^2_n\rangle}$, encoded in
$F_N(q)$. Analyses of these type have been carried out using COHERENT
CsI[Na] data in the limit
$r_\text{rms}^n|_\text{Cs}=r_\text{rms}^n|_\text{I}$, for which
Ref. \cite{Cadeddu:2017etk} found the 1$\sigma$ result
$r_\text{rms}^n|_\text{Cs,I}=5.5^{+0.9}_{-1.1}\,$fm. Later on using the
LAr data release a similar analysis found the $90\%$ CL upper limit
$r_\text{rms}^n<4.33\,$fm \cite{Miranda:2020tif}, a value which mainly
applies to $^{40}$Ar given its natural abundance. Forecasts of neutron
distributions measurements using CE$\nu$NS data have been presented in
Ref. \cite{Coloma:2020nhf}.

In addition to SM measurements, CE$\nu$NS can be used as a probe for
new physics searches. Using COHERENT data, various BSM scenarios have
been studied. They include neutrino non-standard interactions (NSIs) and neutrino generalized interactions, light vector and
scalar mediators interactions, sterile neutrinos and neutrino
electromagnetic properties (see
e.g. \cite{Liao:2017uzy,Coloma:2017ncl,Coloma:2019mbs,Farzan:2018gtr,
  AristizabalSierra:2019ufd,AristizabalSierra:2018eqm,Kosmas:2017tsq,
  Miranda:2019wdy,Papoulias:2019txv,Miranda:2020tif,Dutta:2019eml}).
To illustrate the capabilities of the $\nu$BDX-DRIFT detector and as a
proof of principle, here we focus on NSI scenarios. Given the ingoing
neutrino flavor the couplings that can be proved are $\epsilon_{\mu e}$,
$\epsilon_{\mu\mu}$ and $\epsilon_{\mu\tau}$ (see Section \ref{sec:bsm-cases}
for details). We then focus on these couplings and consider---for
simplicity---a single-parameter analysis.

\begin{figure*}
  \centering
  \includegraphics[scale=0.335]{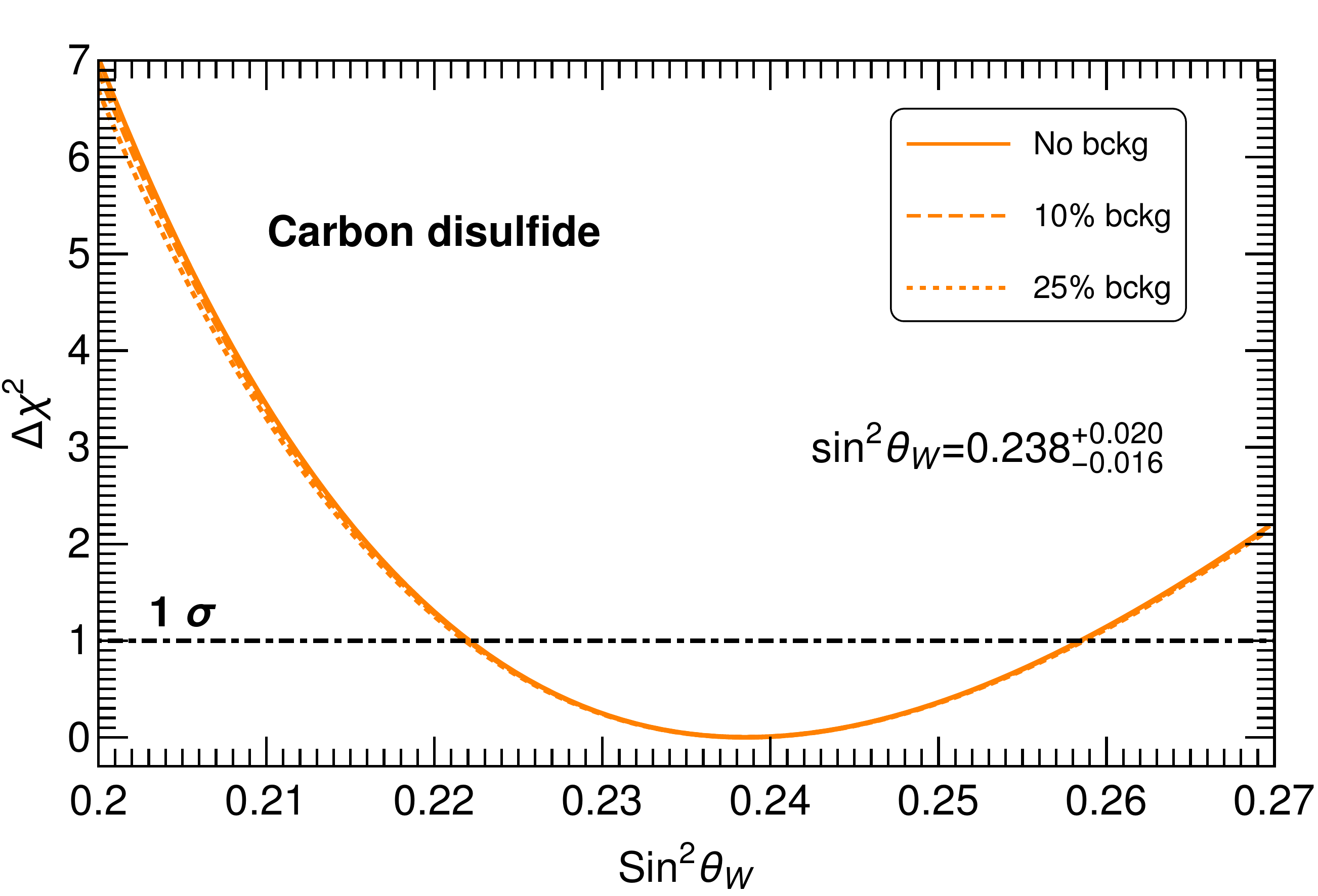}
  \includegraphics[scale=0.335]{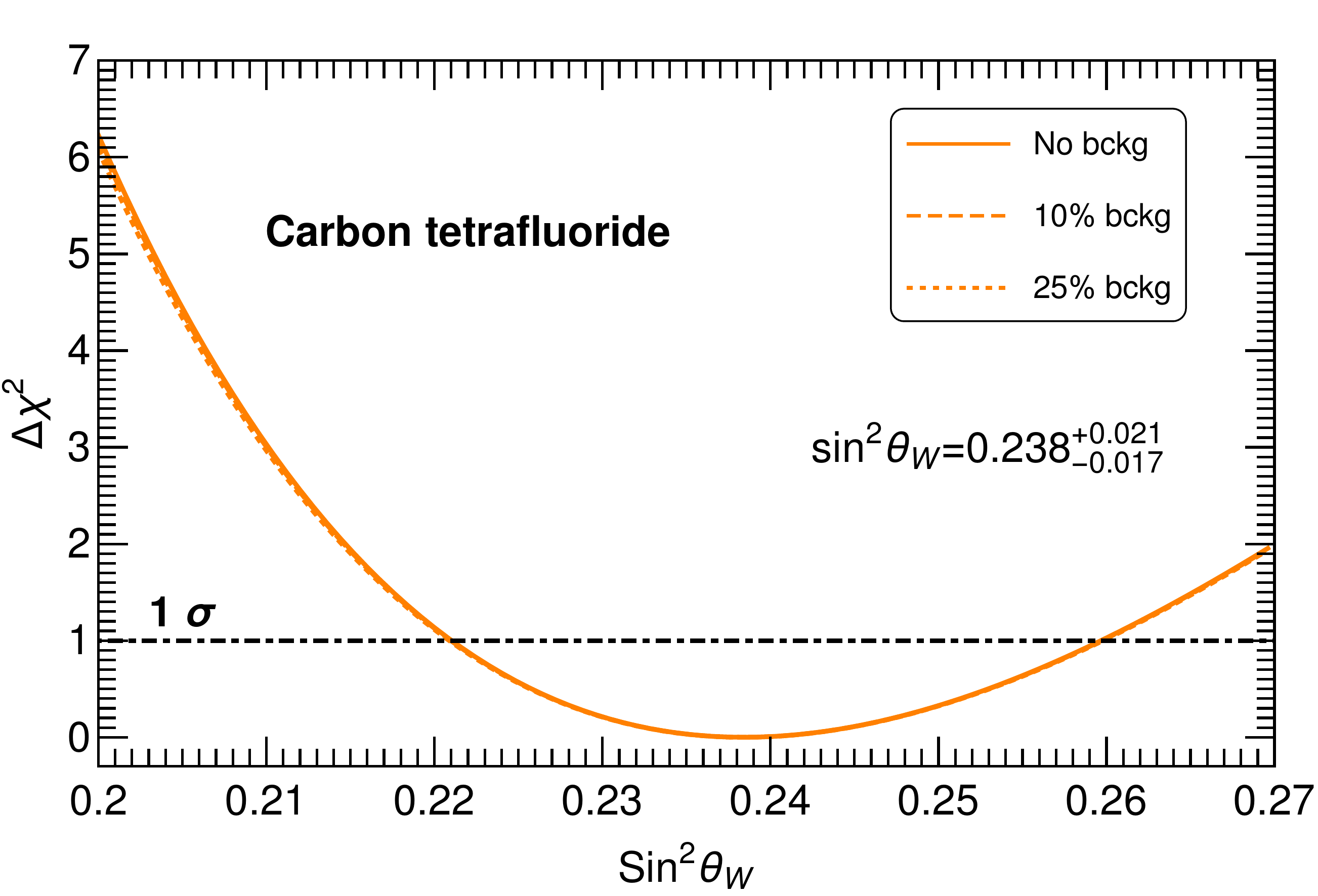}
  \caption{\textbf{Left graph}: Chi-square distribution for
    $\sin^2\theta_W$ assuming a ten-cubic meter detector volume, 7-years data taking and 100\% detector efficiency. The calculation has been done assuming the bulk of the gas is filled with CS$_2$ under three background hypotheses: A free background measurement and $10\%$ and $25\%$ of the measured signal, assumed to be the SM
    prediction at $411\,$Torr with $\sin^2\theta_W$ fixed according to its low-energy
    extrapolation \cite{Kumar:2013yoa}. \textbf{Right graph}: Same as
    left graph but assuming instead that the bulk of the gas is filled with CF$_4$ at $400\,$Torr, pressure at which the SM prediction amounts to 808 events/7-years.}
  \label{fig:weak-mixing-angle-results}
\end{figure*}
We start our discussion with sensitivities of $\nu$BDX-DRIFT to
the weak mixing angle and the rms radii of the neutron distributions for
carbon, fluorine and lead. We then discuss sensitivities to the neutrino
NSI. To determine sensitivities, in all cases we employ a simple
single-bin chi-square analysis with the test statistics defined as \cite{Akimov:2017ade}
\begin{equation}
  \label{eq:chi-square}
  \chi^2=\left(\frac{N_\text{Exp} - (1+\alpha)N_\text{Theo}(p)}
  {\sigma}\right)^2 + \left(\frac{\alpha}{\sigma_\alpha}\right)^2\ ,
\end{equation}
where for $N_\text{Exp}$ we assume the SM prediction adapted to the
case we are interested in (see Sections below), $N_\text{Theo}$
represents predictions of the underlying hypothesis determined by the
values of the parameter(s) 
$p$ and for the statistical
uncertainty we assume $\sigma=\sqrt{N_\text{Exp} + \text{B}}$. Here B
refers to background, which we take to be
$\text{B}=N_\text{Exp}\times f/100$ ($f=0,10,25$). We include as well a
systematic uncertainty $\sigma_\alpha$ along with its nuisance parameter
$\alpha$. In the former we include uncertainties due to the nuclear form
factor $\mathcal{U}_F$ and the neutrino flux $\mathcal{U}_\nu$, which we add
in quadrature. For both we assume $10\%$, see Section
\ref{sec:neutron-density-distributions} and Ref. \cite{Abi:2020evt}.
\subsubsection{Measurements of the weak mixing angle}
\label{sec:weak-mixing-angle}
Measurements of the weak mixing angle not only provide information on
the quantum structure of the SM, but allow indirectly testing new
physics effects at the TeV scale and beyond. The most precise
measurements of $\sin^2\theta_W$ come from: (i) The right-left $Z$
pole production asymmetry measured at SLAC \cite{Abe:2000dq}, (ii) the
$Z\to b\bar b$ forward-backward asymmetry measured at LEP1
\cite{ALEPH:2010aa}. These measurements are known to disagree at the
3.2$\sigma$ level, 
so improved experimental determinations are
required. Low-energy measurements of $\sin^2\theta_W$ aim at doing so,
with different precisions depending on the experimental techniques
employed \cite{Kumar:2013yoa}. Some might be able to reach the level
of precision required, some others may not. However, even those
not reaching that level (order $0.1\%$) will be able to test exotic
contributions to $\sin^2\theta_W$ that could be lurking at low energies.

Low-energy measurements of $\sin^2\theta_W$ at
$\langle q\rangle\ll m_Z\,$ include atomic parity violation in cesium
at $\langle q\rangle\simeq 2.4\,$MeV \cite{Wood:1997zq,Dzuba:2012kx},
electron-electron M{\o}ller scattering at
$\langle q\rangle\simeq 160\,$MeV \cite{Anthony:2005pm}, and
$\nu_\mu$-nucleus deep-inelastic scattering at
$\langle q\rangle\simeq 5\,$GeV \cite{Zeller:2001hh}. More recent
measurements involve electron parity-violating deep-inelastic
scattering at $\langle q\rangle\simeq 6\,$GeV \cite{Wang:2014bba} and
precision measurements of the weak charge of the proton
at $\langle q\rangle\simeq 157\,$MeV \cite{Androic:2018kni}. The precision
of these measurements range from $\pm 0.4\%$ for the weak charge of
the proton up to $\pm 4\%$, for electron parity-violating
deep-inelastic scattering. Thus, none of them have the level of precision
achieved at $Z$ pole measurements, but are precise enough to
constraint new physics effects. Future atomic parity violation
experiments as well as ultraprecise measurements of parity violation in
electron-$^{12}$C scattering will improve the determination of
$\sin^2\theta_W$ at the $\sim 0.1\%$ level \cite{Kumar:2013yoa}.

As it has been already stressed, CE$\nu$NS provides another
experimental environment in which information on $\sin^2\theta_W$ can
be obtained. Probably the most ambitious scenario is that of reactor
neutrinos: the combination of a large neutrino flux and small baseline
provides large statistics with which the weak mixing angle can be
determined with a precision of $\pm 0.1\%$ or even below,
depending on detector efficiency and systematic errors
\cite{Canas:2018rng}. For spallation neutron source neutrinos, current
precision is of order $\pm 50\%$. However, expectations are that data from
future ton-size detectors (LAr and NaI[Tl]) will improve this measurement.

To assess the precision at which $\nu$BDX-DRIFT can measure the weak
mixing angle we assume two detector configurations in which the bulk
of the gas is filled with either carbon disulfide or carbon tetrafluoride.
For CS$_2$ we take the detector pressure to be $411\,$Torr,
while for CF$_4$ $400\,$Torr. In both cases a $100\%$ detector
efficiency is assumed. For $N_\text{exp}$ we assume the SM prediction calculated
with $\sin^2\theta_W$ extrapolated to low energies \cite{Kumar:2013yoa}:
\begin{equation}
  \label{eq:weak-mix-angle-extrapolated}
  \sin^2\theta_W(q=0)=\kappa(q=0)_{\overline{\text{MS}}}
  \sin^2\theta_W(m_Z)_{\overline{\text{MS}}}\ ,
\end{equation}
with $\kappa(q=0)_{\overline{\text{MS}}}=1.03232\pm 0.00029$ and
$\sin^2\theta_W(m_Z)_{\overline{\text{MS}}}=0.23122\pm 0.00003$
\cite{Tanabashi:2018oca}. For the calculation we take only central values.
With the toy experiment fixed, we then calculate $N_\text{Theo}$ for
$\sin^2\theta_W\subset [0.20,0.27]$, for which we find that the event
yield varies from $280$ to $507$ events for carbon disulfide and
from $589$ to $1012$ events for carbon tetrafluoride.

The results of the chi-square analysis are shown in
Fig. \ref{fig:weak-mixing-angle-results}, left graph for carbon disulfide
and right graph for carbon tetrafluoride. The level at which
$\sin^2\theta_W$ can be determined depends---of course---on the amount
of background, although its impact is not severe. Assuming the detector
is operated under zero background conditions we get for both CS$_2$ and
CF$_4$ the $1\sigma$ results:
\begin{alignat}{2}
  \text{CS$_2$:}&\quad\sin^2\theta_W&=0.238^{+0.020}_{-0.016}\ ,
  \nonumber\\
  \text{CF$_4$:}&\quad\sin^2\theta_W&=0.238^{+0.021}_{-0.017}\ .
\end{alignat}
From these results one can see that the precision with which the weak mixing
angle can be measured at $\nu$BDX-DRIFT will be of order $8\%$. That
precision exceeds what has been so far achieved with any of the COHERENT
detectors, and comparable to what DUNE 7-years data taking could
achieve in the electron recoil channel, $3\%$.

\begin{figure*}[t]
  \centering
  \includegraphics[scale=0.337]{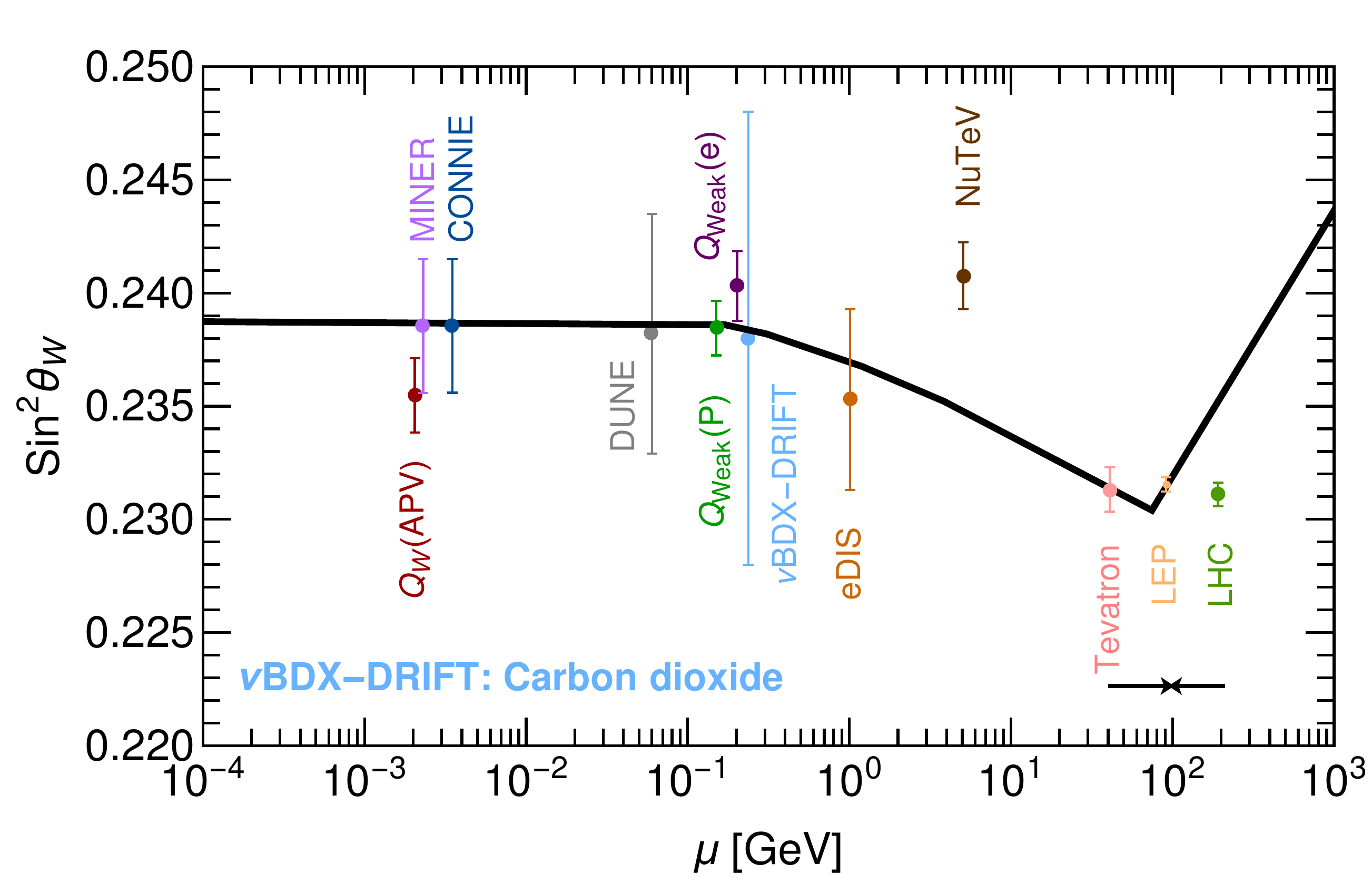}
  \includegraphics[scale=0.33]{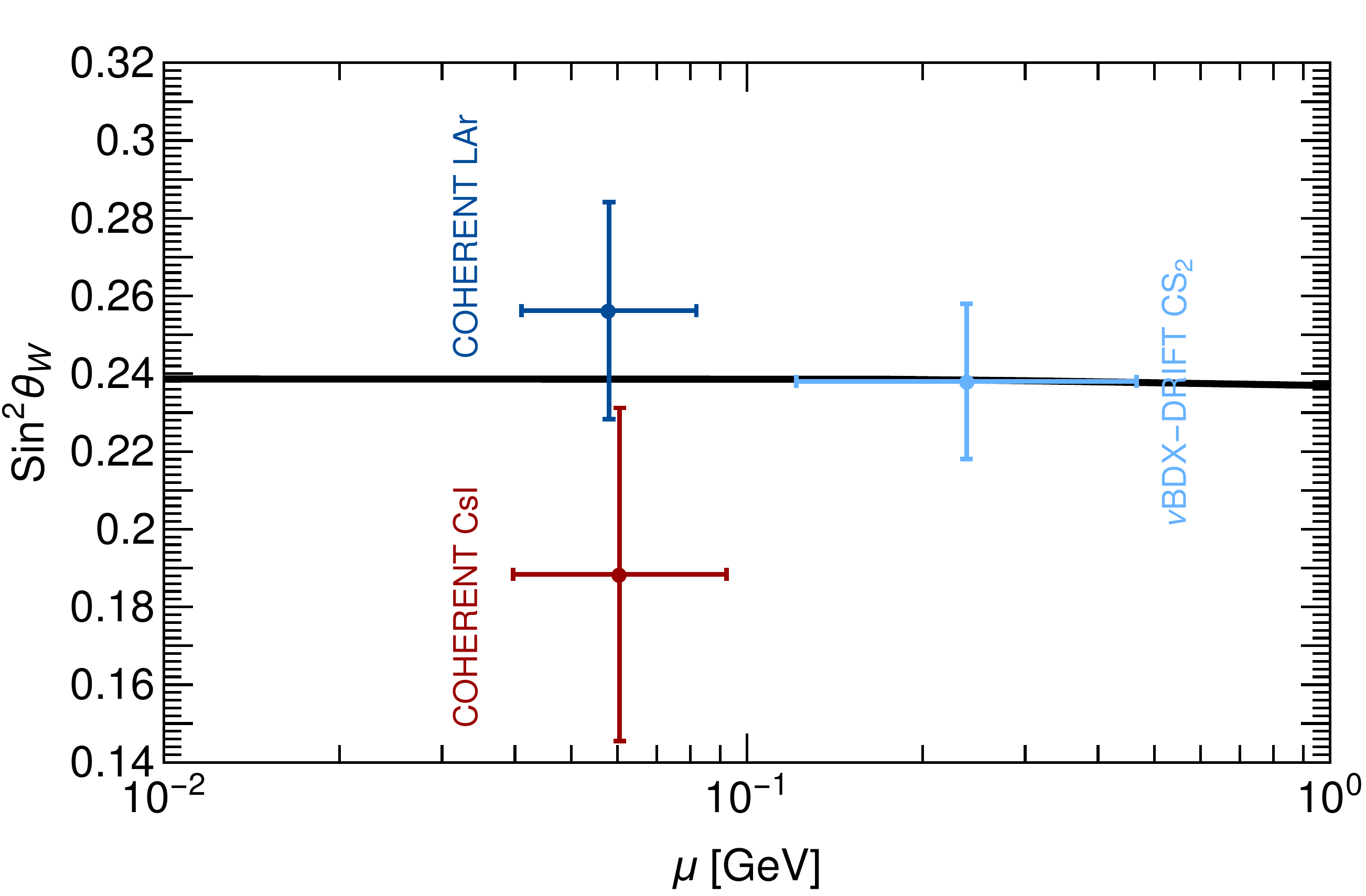}
  \caption{\textbf{Left graph}: Weak mixing angle RGE evolution in the
    $\overline{\text{MS}}$ renormalization scheme as calculated in
    Ref. \cite{Erler:2004in}, along with a variety of measurements at
    different renormalization scales: Atomic parity violation (APV)
    \cite{Wood:1997zq,Dzuba:2012kx}, MINER
    \cite{Agnolet:2016zir,Canas:2018rng}, CONNIE
    \cite{Aguilar-Arevalo:2019jlr,Canas:2018rng} (slightly offset
    horizontally for clarity), proton weak charge ($Q_\text{Weak}(P)$)
    from cesium transitions \cite{Androic:2018kni}, electron weak
    charge ($Q_\text{Weak}(e)$) from M{\o}ller scattering
    \cite{Anthony:2005pm}, parity violation in electron deep inelastic
    scattering (eDIS) \cite{Wang:2014bba}, neutrino-nucleus scattering
    (NuTeV) \cite{Zeller:2001hh} and collider measurements (Tevatron, LEP and LHC. LEP and LHC results offset horizontally as indicated by the arrows)
    \cite{Tanabashi:2018oca}. Results for the DUNE near
    detector using elastic neutrino-electron scattering are displayed
    as well \cite{deGouvea:2019wav}. The result for $\nu$BDX-DRIFT follows
    from the chi-square analysis in the left graph in
    Fig. \ref{fig:weak-mixing-angle-results} and the error bar has
    been reduced by a factor 2 to allow comparison with the other
    experiments. \textbf{Right graph}: Same as left graph but for
    fixed-target CE$\nu$NS experiments, COHERENT CsI[Na]
    \cite{Akimov:2017ade,Kosmas:2017tsq} and LAr
    \cite{Akimov:2020czh,Miranda:2020tif}.  This result shows that
    measurements at $\nu$BDX-DRIFT can be more competitive that those
    obtained so far with COHERENT data, thus providing complementary information in the nuclear recoil channel to DUNE near detector measurements using the electron channel instead.}
  \label{fig:rge-weak-mix-angle}
\end{figure*}
To put in perspective the precision that can be achieved at
$\nu$BDX-DRIFT, we have plotted the RGE evolution of the weak mixing
angle in the $\overline{\text{MS}}$ renormalization scheme along with
the low-energy measurements of the high precision experiments we have
discussed. We have as well included expectations from the DUNE near
detector using elastic neutrino-electron scattering
\cite{deGouvea:2019wav}. The result is shown in Fig. \ref{fig:rge-weak-mix-angle},
left graph. To allow comparison we have reduced the error bar by a factor $2$.
One can see that although $\nu$BDX-DRIFT  comes with a larger uncertainty than these
high-precision experiments, it brings information at a renormalization
scale which is not covered by any of those experiments. We note that the precise location of the scale constrained by the experiment depends on detectors parameters such as the assumed recoil threshold and the shape of the neutrino spectrum. In Fig.~\ref{fig:rge-weak-mix-angle} we simply plot it at the scale corresponding to the mean recoil energy, which we find agrees within uncertainty with a more rigorous calculation accounting for the shape of the neutrino spectrum. Note that the result we obtain is
 expected, as it is known that reaching order $\pm 1\%$ precision
in neutrino scattering experiments is challenging \cite{Kumar:2013yoa}.

Note that if one focuses on experiments that fall within the same
``category'' (stopped-pion CE$\nu$NS-related experiments) then a more
reliable comparison can be done. The right graph in Fig. \ref{fig:rge-weak-mix-angle}
shows the $1\sigma$ sensitivities for COHERENT CsI[Na] and LAr along with
what can be achieved at $\nu$BDX-DRIFT. We have included as well the 
$\mu=\langle q\rangle$ range that these experiments cover. For that we have
used $q^2=2m_NE_r$ along with information on the minimum and maximum
recoil energies these experiments have measured, or will in the case
of $\nu$BDX-DRIFT: COHERENT CsI[Na], $E_r\subset [5,30]\,\text{keV}$
\cite{Akimov:2017ade}; COHERENT LAr, $E_r\subset [19,81]\,\text{keV}$
\cite{Akimov:2020czh}; $\nu$BDX-DRIFT CS$_2$, $E_r\subset [101,2640]\,\text{keV}$.
For the latter we have used $E_\nu^\text{min}=39\,$MeV and
$E_\nu^\text{min}=200\,$MeV, values dictated by the neutrino spectrum low-energy
tail and the coherence condition. These values result in
\begin{align}
    \text{CsI}:&\quad q\subset [35,86]\times 10^{-3}\,\text{GeV}\ ,
    \nonumber\\
    \text{LAr}:&\quad q\subset [38,78]\times 10^{-3}\,\text{GeV}\ ,
    \nonumber\\
    \text{CS}_2:&\quad q\subset [78,397]\times 10^{-3}\,\text{GeV}\ ,
\end{align}
and $\langle q\rangle=61\times 10^{-3}\,$ GeV, $\langle q\rangle=58\times 10^{-3}\,$
GeV and $\langle q\rangle=238\times 10^{-3}\,$ GeV, respectively. One can see that
among those stopped-pion CE$\nu$NS experiments $\nu$BDX-DRIFT has a better performance.
\begin{figure*}
    \centering
    \includegraphics[scale=0.33]{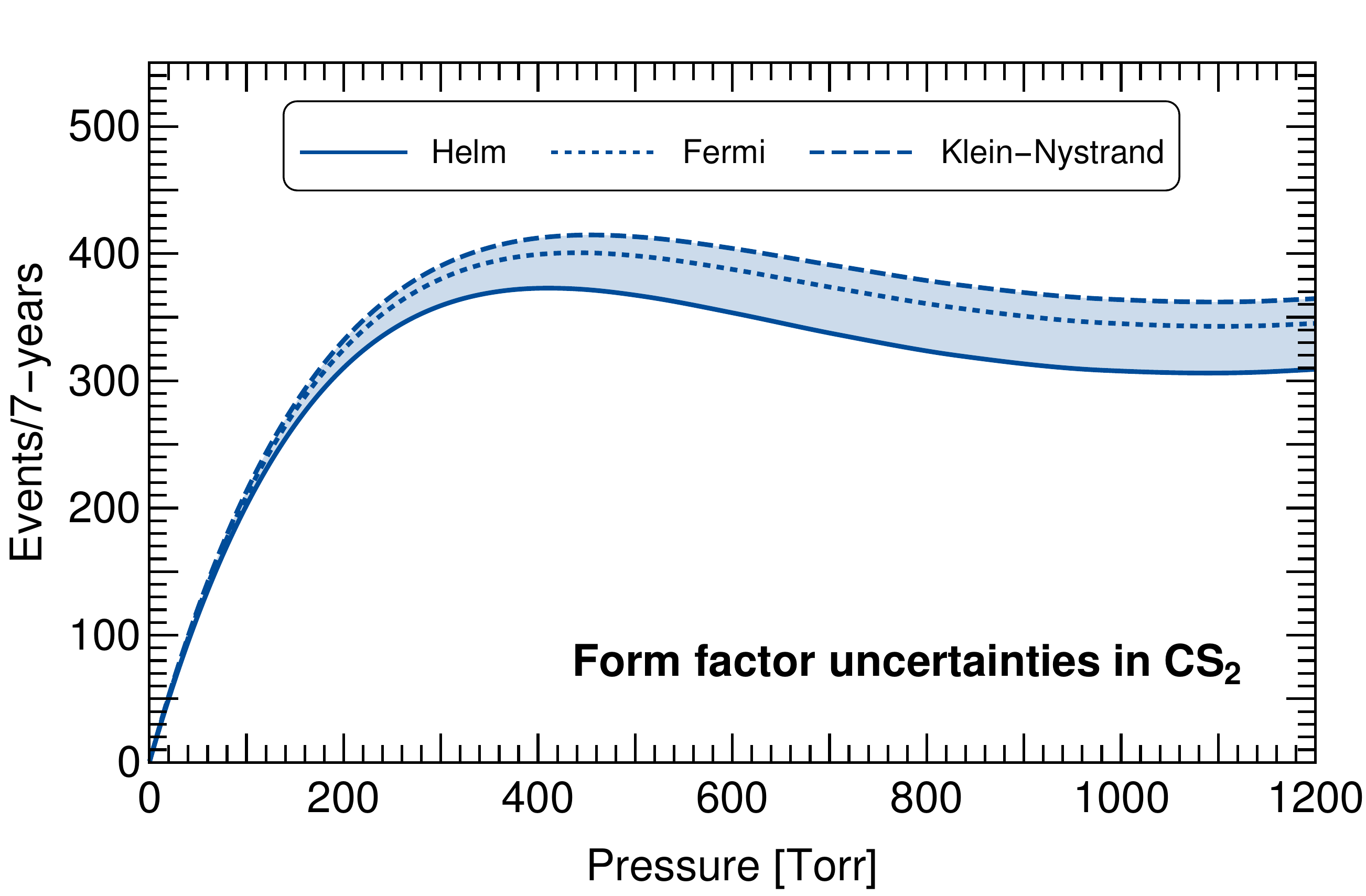}
        \includegraphics[scale=0.33]{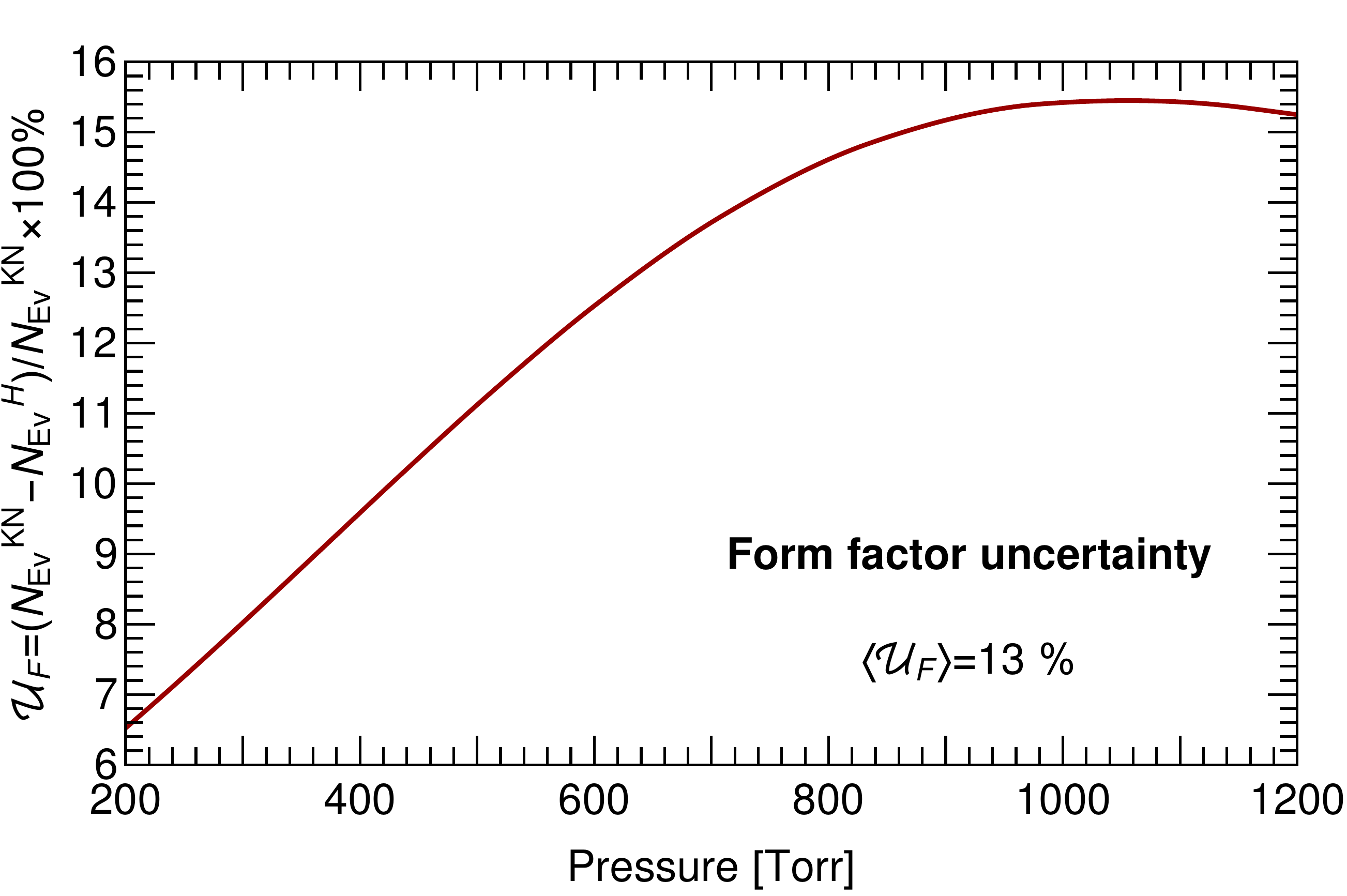}
    \caption{\textbf{Left graph}: CE$\nu$NS event yield as a function of
        pressure for carbon dioxide assuming a ten-cubic meter detector volume and 7-years
        data taking. The calculation has been done assuming three
        different form factor parametrizations: Helm form factor, symmetrized Fermi form factor
        and Klein-Nystrand form factor. The result shows that the smallest (largest) event yield
        is obtained from the Helm (Klein-Nystrand) form factor. \textbf{Right graph}:
        Percentage uncertainty as a function of pressure calculated from the minimum and
        maximum event yields. As pressure increases the difference increases as well as a
        result of increasing neutrino energy.}
    \label{fig:form-factor-uncertainties}
\end{figure*}
\subsubsection{Form factor uncertainties and measurements of neutron density distributions}
\label{sec:neutron-density-distributions}
Given the recoil energies involved in the $\nu$BDX-DRIFT experiment, one
expects the CE$\nu$NS event yield to be rather sensitive to nuclear physics effects.
Thus to assess the degree at which these effects affect CE$\nu$NS predictions, we
first calculate the intrinsic uncertainties due to the form factor parametrization
choice. For that aim we use---in addition to the Helm form factor parametrization
\cite{Helm:1956zz}---the Fourier transform of the symmetrized Fermi distribution
and the Klein-Nystrand form factor \cite{0305-4470-30-18-026,Klein:1999qj}.

The Helm model assumes that the proton and neutron distributions are dictated by
a convolution of a uniform density of radius $R_0$ and a Gaussian profile
characterized by the folding width $s$, responsible for the surface thickness. The
Helm form factor then reads \cite{Helm:1956zz}
\begin{equation}
    \label{eq:Helm-FF}
    F_\text{H}(q^2)=3\frac{j_1(qR_0)}{qR_0}e^{-(qs)^2/2}\ ,
\end{equation}
where $j_1$ is the spherical Bessel function of order one and $R_0$, the diffraction
radius, is determined by the surface thickness and the rms radius of the corresponding
distribution, namely \cite{Lewin:1995rx}
\begin{equation}
    \label{eq:diffraction-radius-Helm}
    R_0=\sqrt{\frac{5}{3}\left(\langle r^2\rangle_\text{H} - 3s^2\right)}\ .
\end{equation}
For the surface thickness we use $0.5\,$fm \cite{Lewin:1995rx}. The symmetrized Fermi
form factor follows instead from the symmetrized Fermi function, defined through the
conventional Fermi or Woods-Saxon function. The resulting form factor is given by
\cite{0305-4470-30-18-026}
\begin{align}
    \label{eq:symmetrized-Fermi-FF}
    F_\text{SF}(q^2)=&\frac{3}{qc}\left[\frac{\sin(qc)}{(qc)^2}
    \left(\frac{\pi qa}{\tanh(\pi qa)}-\frac{\cos(qc)}{qc}\right)\right]
    \nonumber\\
    &\frac{\pi qa}{\sinh(\pi qa)}\frac{1}{1+(\pi a/c)^2}\ .
\end{align}
Here $c$ defines the half-density radius and $a$ the surface diffuseness, both 
related through the rms radius of the distribution
\begin{equation}
    \label{eq:half-density-radius-Fermi}
    c=\sqrt{\frac{5}{3}\left(\langle r^2\rangle_\text{SF}
    -\frac{7}{5}(\pi a)^2\right)}\ .
\end{equation}
For the calculation we fix $a=0.52\,$fm \cite{Piekarewicz:2016vbn}. Results are
rather insensitive to reasonable changes of this parameter
\cite{AristizabalSierra:2019ufd}. Finally, the Klein-Nystrand form factor follows from
folding a Yukawa potential of range $a_k$ over a hard sphere distribution with
radius $R_A$. The form factor is then given by \cite{Klein:1999qj}
\begin{equation}
    \label{eq:KN-FF}
    F_\text{KN}=3\frac{j_1(qR_A)}{qR_A}\frac{1}{1+q^2a_k^2}\ .
\end{equation}
In this case the radius $R_A$ and the potential range $a_k$ are related through the rms
radius of distribution according to
\begin{equation}
    \label{eq:radius-potential-range-KN-FF}
    R_A=\sqrt{\frac{5}{3}\left(\langle r^2\rangle_\text{KN} - 6a_k\right)}\ ,
\end{equation}
with the value for $a_k$ given by 0.7 fm \cite{Klein:1999qj}.

With these results at hand we are now in a position to calculate the CE$\nu$NS event yield.
We do so for carbon disulfide assuming the detector specifications used in our previous
analyses. The result is displayed in Fig. \ref{fig:form-factor-uncertainties} left graph, from which it can be seen that the event yield has a relative mild dependence on the nuclear form factor choice. The minimum and maximum values interpolate between the results obtained using the Helm and Klein-Nystrand form factors. It is worth noting that for reactor neutrinos, form
factor effects are completely neglible while for SNS neutrinos (COHERENT) they are mild,
of order $5\%$ or so \cite{Akimov:2017ade}. In this case the dependence is stronger,
a result expected given the energy regime of the neutrino probe. Right graph in 
Fig. \ref{fig:form-factor-uncertainties} shows the percentage uncertainty calculated
according to
\begin{equation}
    \label{eq:percentage-uncertainty}
    \mathcal{U}_F=\frac{N_\text{ev}^\text{KN}-N_\text{ev}^\text{H}}
        {N_\text{ev}^\text{KN}}\times 100\%\ ,
\end{equation}
and covering pressures up to 1200 Torr. From this result
it can be seen that at low recoil energy thresholds uncertainties are of order $6-7\%$, and raise
up to order $16\%$ at high recoil energy thresholds. Calculation of the average uncertainty results in $\langle\mathcal{U}_F\rangle\simeq 13\%$. This means that calculation of SM CE$\nu$NS
predictions as well as possible new physics effects always come along with such
uncertainty. Note that our calculations in the previous Sections, based on the
Helm form factor, should be understood as lower limit predictions of what should be expected.

\begin{figure*}
    \centering
    \includegraphics[scale=0.335]{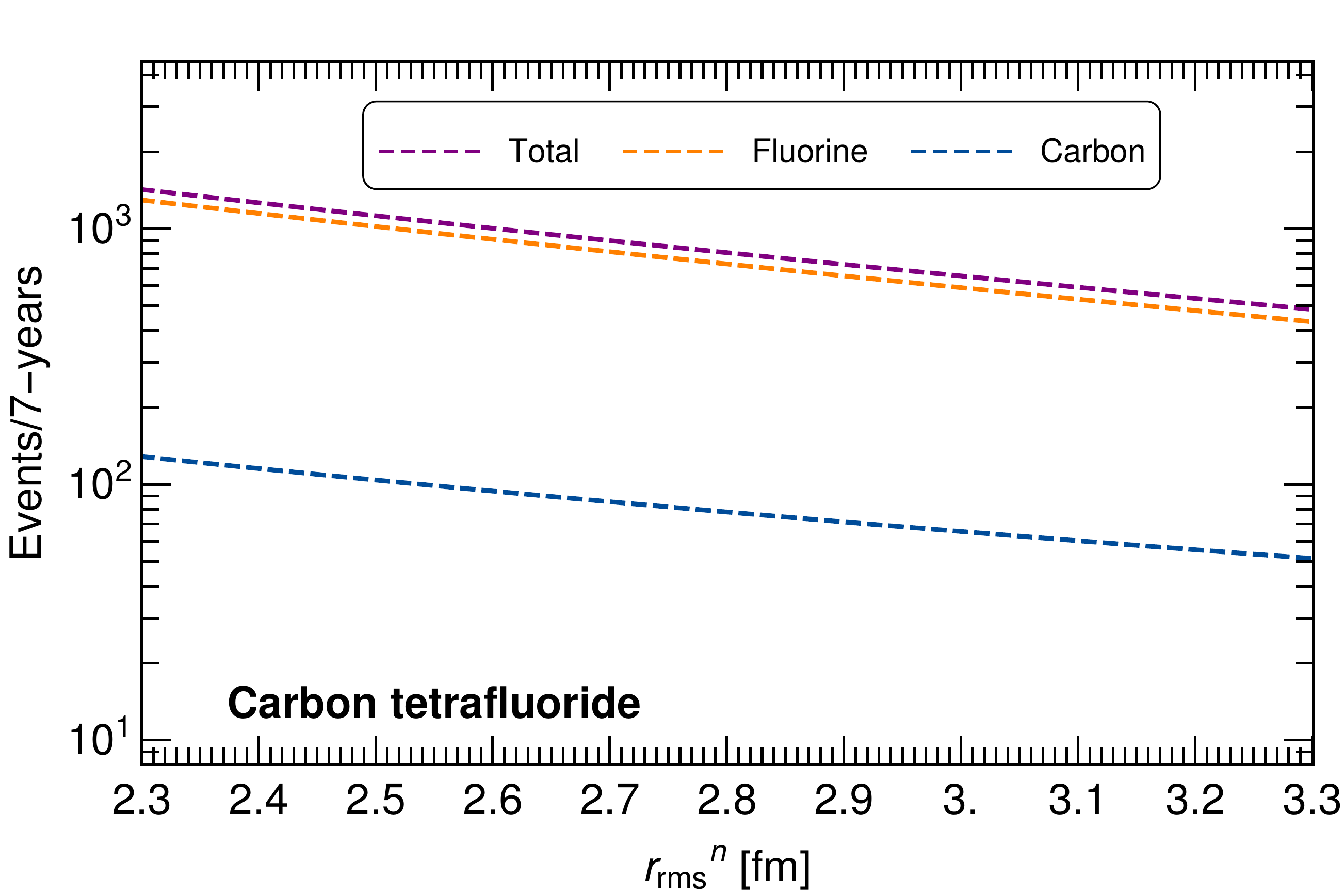}
    \includegraphics[scale=0.33]{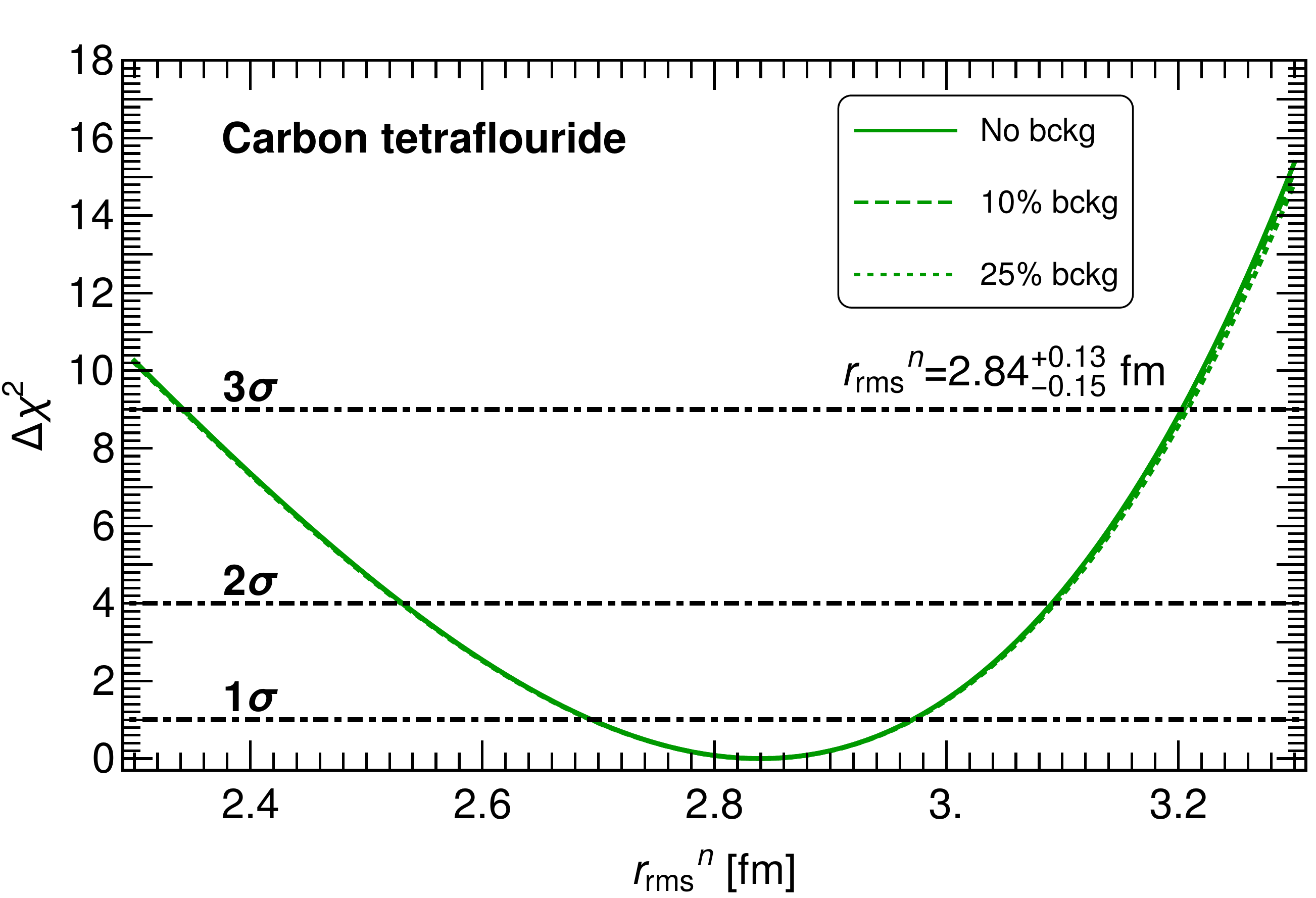}
    \includegraphics[scale=0.335]{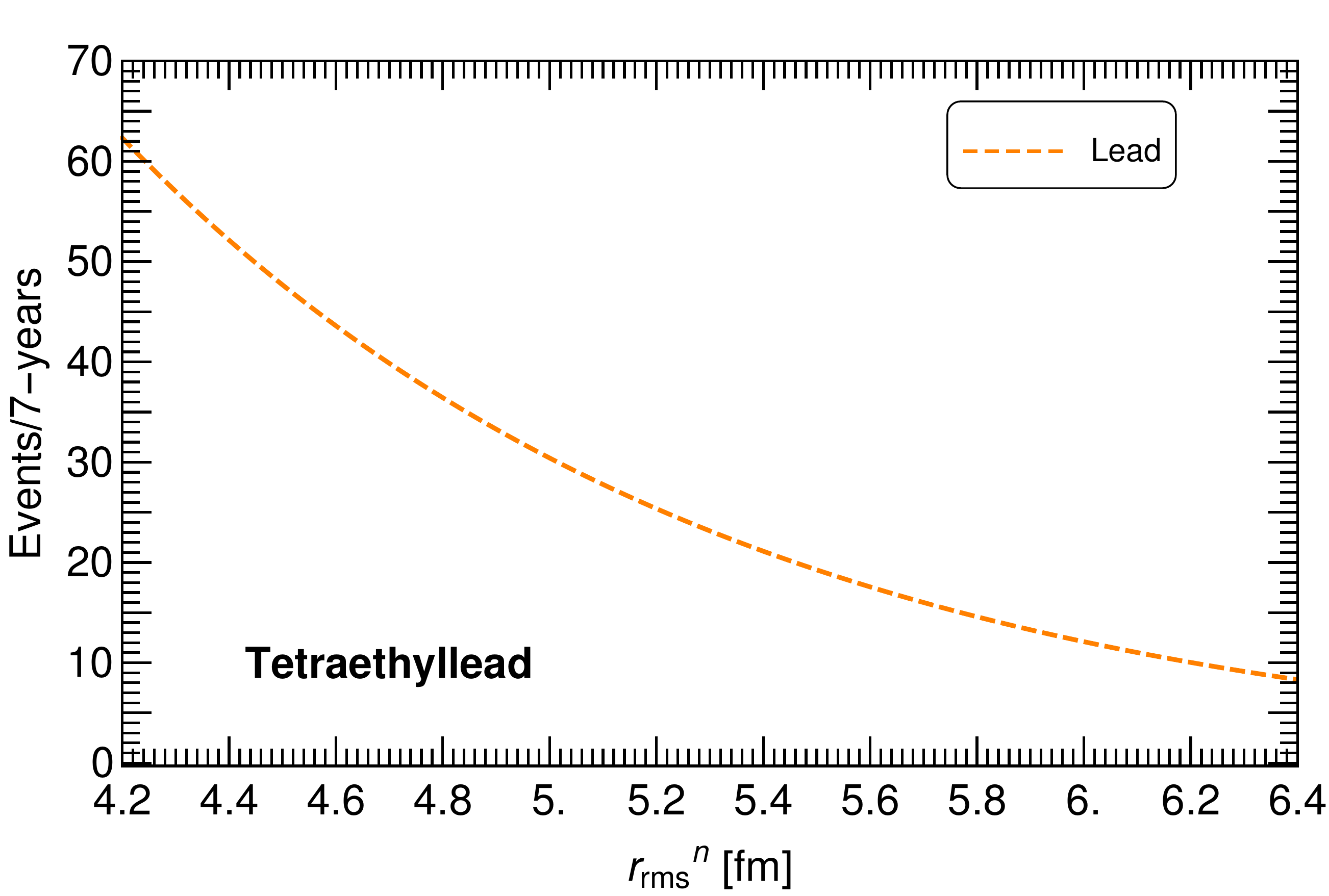}
    \includegraphics[scale=0.33]{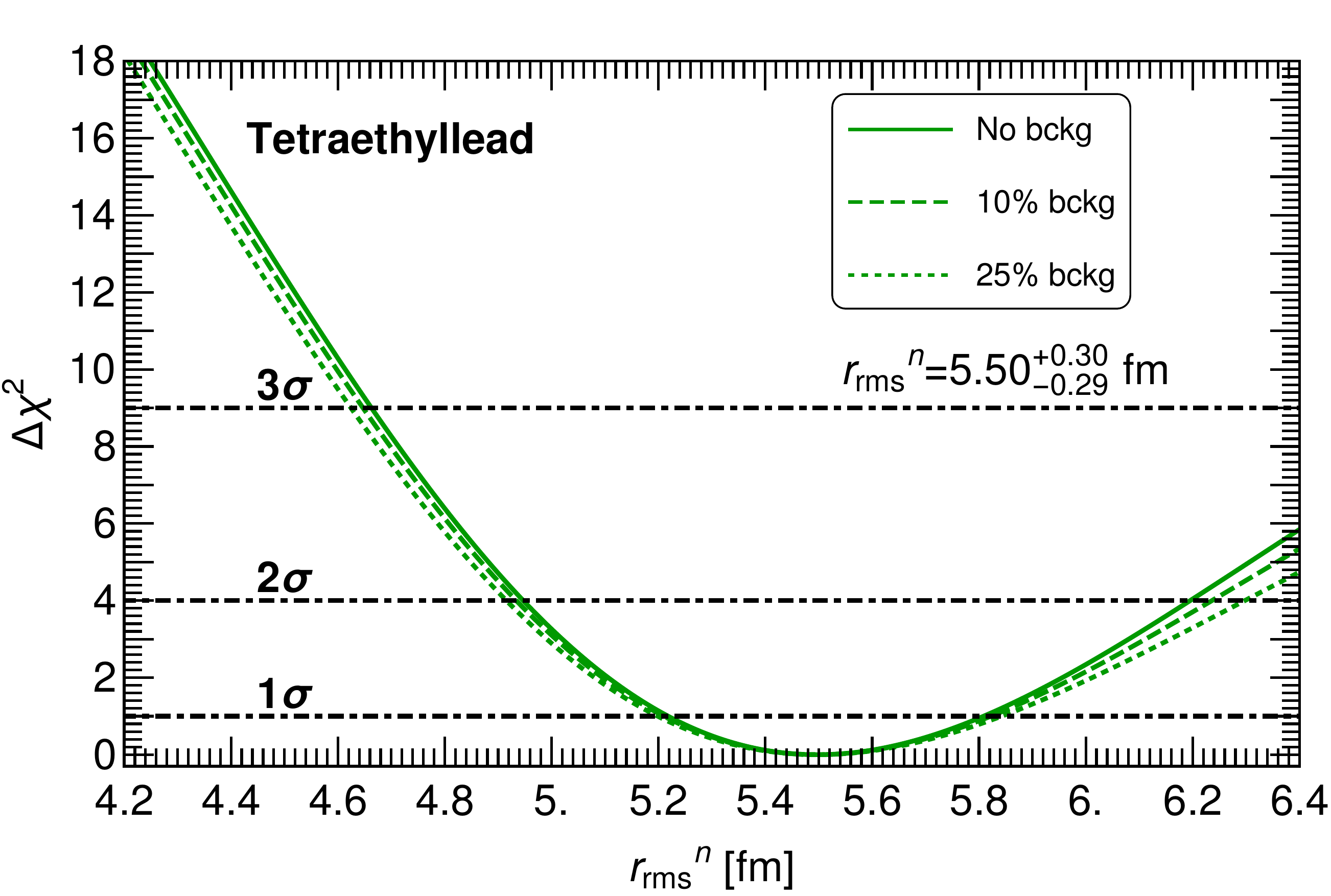}
    \caption{\textbf{Top-left graph}: Event distribution in terms of the neutron rms radius
    for $\text{C}\text{F}_4$. \textbf{Top-right graph}: Chi-square distribution for
    the neutron rms radii of carbon and fluorine including our three background hypotheses.
    For the calculation we have assumed a ten-cubic meter detector volume, 7-years data taking
    and 100\% detector efficiency. The value for the neutron rms radius follows from the
    background-free case, a potential experimental scenario given the directional properties
    of the $\nu$BDX-DRIFT detector. We assume 100\% of the detector is filled with
    $\text{C}\text{F}_4$. \textbf{Bottom-left graph}: Event distribution in terms of the
    neutron rms radius for $\text{C}_8\text{H}_{20}\text{Pb}$. \textbf{Bottom-right graph}:
    Chi-square distribution for the neutron rms radius of lead under the assumptions used
    in the carbon and fluorine case. A 2.3:1 ($\text{C}\text{S}_2:\text{C}_8\text{H}_{20}\text{Pb}$)
    gas ratio has been assumed. The chi-square analyses include systematics
    due to form factor parametrization dependences as well as neutrino flux uncertainties.}
    \label{fig:rms_of_neutron_distributions}
\end{figure*}
We now turn to the discussion of measurements of neutron distributions, in particular, of the rms radius of the neutron distribution. This quantity is relevant since, combined 
with the rms radius of the proton distribution, it defines the neutron skin thickness of
a nucleus, $\Delta r_{np}(\text{nucleus})=r_\text{rms}^n - r_\text{rms}^p$.
This quantity in turn is relevant in nuclear physics as well as in astrophysics.
For instance, in nuclear physics it plays an important role in the nuclear energy density
functional \cite{Reinhard:2010wz,Reinhard:2013fpa,Nazarewicz:2013gda,Chen:2014mza,Chen:2014sca},
while in astrophysics it allows the prediction of neutron star properties such as its density
and radius \cite{Horowitz:2000xj}.

A clean direct measurement of the neutron rms radius has been done only for $^{208}\text{Pb}$
by the PREX experiment at the Jefferson laboratory \cite{Abrahamyan:2012gp,Horowitz:2013wha}.
The rms radii for other nuclides have been mapped using hadronic experiments, and suffer
from large uncontrolled uncertainties \cite{Horowitz:2014bja}. In contrast to these experiments, PREX relies
on parity-violating elastic electron scattering thus providing a clean determination not only
of the neutron rms radius but of the neutron skin of $^{208}\text{Pb}$. As we have already mentioned,
CE$\nu$NS experiments provide an alternative experimental avenue to determine
this quantity for other nuclides. 

To prove the capabilities of the $\nu$BDX-DRIFT detector we calculate
sensitivities for carbon, fluorine and lead. Measurements of the rms radius of the
neutron distribution for carbon and fluorine can be done using CF$_4$. Since carbon
and fluorine have about the same amount of neutrons, in first approximation one
can assume $r_\text{rms}^n|_\text{C}=r_\text{rms}^n|_\text{F}=r_\text{rms}^n$. 
The analysis for lead can be done using instead $\text{C}_8\text{H}_{20}\text{Pb}$.
In this case the large mismatch between the number of neutrons for carbon and
lead does not allow the approximation employed for $\text{C}\text{F}_4$.
Experimentally, however, that measurement could be carried out by tuning pressure
to the value at which the lead signal peaks (6.4 Torr) and then selecting lead events.
The latter enabled by the different ranges for carbon and lead given an ionization.
Following this strategy we then calculate $r_\text{rms}^n|_\text{Pb}$ using only the
lead signal. Note that this analysis intrinsically assumes that all lead stable
nuclei ($^{204}$Pb, $^{206}$Pb, $^{207}$Pb and $^{208}$Pb) have the same
$r_\text{rms}^n$. This of course is not the case, but it is a
rather reasonble assumption given the precision at which the $r_\text{rms}^n$ can be
measured at $\nu$BDX-DRIFT.

To determine sensitivities we use as toy experiment input the SM prediction assuming
$r_\text{rms}^n=\langle r_\text{rms}^p\rangle$, where $\langle r_\text{rms}^p\rangle$
is calculated according to $\sum_i r_{\text{rms}_i}^p X_i$ with $r_{\text{rms}_i}^p$
the proton rms radius of $i^\text{th}$ stable isotope \cite{Angeli:2013epw} and $X_i$
its natural abundance. We then perform our statistical analysis by calculating the event
yield by varying $r_\text{rms}^n$ within $[2.3,3.3]\,$fm for CF$_4$ and $[4.2,6.4]\,$fm
for $\text{C}_8\text{H}_{20}\text{Pb}$. Results are shown in Fig. \ref{fig:rms_of_neutron_distributions}. Top-left and bottom-left graphs show the
variation of the event rate in terms of $r_\text{rms}^n$ for $\text{C}\text{F}_4$ and 
$\text{C}_8\text{H}_{20}\text{Pb}$ respectively. One can see that the signal increases
with decreasing $r_\text{rms}^n$, a behavior that can be readily understood from the
reduction in nuclear size implied by a smaller $r_\text{rms}^n$: As nuclear size reduces, coherence extends to larger transferred momentum.

Results of the chi-square analyses are shown in the top-right and bottom-right graphs.
In each case results for our three background hypotheses are displayed. These results
demonstrate that the ten-cubic meter and 7-years data taking $\nu$BDX-DRIFT will be
able to set the following $1\,\sigma$ measurements:
\begin{alignat}{2}
    \text{C and F in CF$_4$}:&\quad r_\text{rms}^n=2.84^{+0.1 3}_{-0.15}\;\text{fm}\ ,
    \nonumber\\
    \text{Pb in C$_8$H$_{20}$Pb}:&\quad r_\text{rms}^n=5.50^{+0.30}_{-0.29}\;\text{fm}\ .
\end{alignat}
From these numbers one can see that the neutron rms radius for carbon and fluorine can be determined at the $3\%$ accuracy level, while for lead at about $5\%$. The difference in precision has to do with the difference in statistics. For $\text{C}\text{F}_4$ about 800 events are available, while for lead in
$\text{C}_8\text{H}_{20}\text{Pb}$ only about 19 due to the constraints implied by the differentiation between lead and carbon events. Note that these measurements will 
not only provide information on these quantities, but can potentially be used to improve 
attempts to reliably extract neutron star radii, in particular those for lead.
\begin{table*}
    \setlength{\tabcolsep}{20pt}
    \renewcommand{\arraystretch}{1.8}
    \centering
    \begin{tabular}{|c||c||c||c|}\hline
    \multicolumn{2}{|c||}{\textbf{$\boldsymbol{\nu}$BDX-DRIFT CS$_2$} $\quad$(7-years)} 
    & \multicolumn{2}{|c|}{\textbf{COHERENT CsI} $\quad$(1-year)}\\\hline
    $\epsilon_{\mu\mu}^u$   & $[-0.013,0.011]\oplus [0.30,0.32]$ 
    & $\epsilon_{\mu\mu}^u$ & $[-0.06,0.03]\oplus [0.37,0.44]$ \\\hline
    $\epsilon_{e\mu}^u$     & $[-0.064,0.064]$ & $\epsilon_{e\mu}^u$ & $[-0.13,0.13]$\\\hline
    \end{tabular}
    \caption{$1\sigma$ allowed ranges for neutrino NSI couplings derived from a single-parameter analysis. Results for down quark parameters are rather close to those derived for up quarks, so are not displayed. Intervals for $\epsilon_{\mu\tau}^q$ ($q=u,d$) are identical to those for $\epsilon_{e\mu}^q$. For 1-year data taking sensitivities can be degraded by up to a factor 5. Allowed $1\sigma$ limits from COHERENT CsI including spectra and timing information are taken from Ref. \cite{Giunti:2019xpr}, are shown for comparison.}
    \label{tab:NSI-summary}
\end{table*}
\subsubsection{Sensitivities to neutrino NSI}
\label{sec:bsm-cases}
Neutrino NSI are four-fermion contact interactions which parametrize a new vector force 
relative to the electroweak interaction in terms of a set of twelve flavored-dependent
new parameters (in the absence of CP-violating phases). Explicitly they read
\cite{Wolfenstein:1977ue}
\begin{equation}
    \mathcal{L}_\text{NSI}=-\sqrt{2}G_F\sum_{q=u,d}
    \overline{\nu}_a\gamma_\mu(1-\gamma_5)\nu_b
    \overline{q}\gamma^\mu\left(\epsilon_{ab}^{Vq}+\epsilon_{ab}^{Aq}\gamma_5\right)q\ ,
\end{equation}
where $a,b\cdots $ are lepton flavor indices. The axial current parameters generate
spin-dependent interactions and hence are poorly constrained. For that 
reason most NSI analyses consider only vector couplings $\epsilon_{ab}^q\equiv\epsilon^{Vq}_{ab}$.
Limits on NSI are abundant and follow from a variety of measurements which include neutrino
oscillations experiments \cite{Escrihuela:2011cf,Gonzalez-Garcia:2015qrr},
low energy scattering processes \cite{Coloma:2017egw} and LHC data \cite{Friedland:2011za,Franzosi:2015wha}.
In the light of COHERENT CE$\nu$NS data they have been extensively considered as well \cite{Liao:2017uzy,Kosmas:2017tsq,Miranda:2020tif,Giunti:2019xpr,Dutta:2020che}, and their potential
experimental traces have been the subject of studies in multi-ton DM experiments \cite{Dutta:2017nht,AristizabalSierra:2017joc,Gonzalez-Garcia:2018dep,AristizabalSierra:2019ykk}.

The presence of neutrino NSI modify the CE$\nu$NS differential cross section. Being vector 
interactions the flavor-diagonal couplings interfere with the SM contribution,
that interference can be constructive or destructive depending on the sign the coupling
comes along with. In contrast, off-diagonal couplings always enhance the SM cross section.
Asssuming equal rms radii for the proton and neutron distributions, the modified cross section
proceeds from Eq. (\ref{eq:x-sec-cevns}) by changing the coherent weak charge according to
\cite{Barranco:2005yy}
\begin{align}
    \label{eq:NSI-change-xsec}
    Q_{Wa}^2&=\left[Z(g_V^p + 2\epsilon_{aa} + \epsilon_{aa}^d) 
    + (A-Z)(g_V^n + \epsilon_{aa} + \epsilon_{aa}^d)\right]^2
    \nonumber\\
    &+\sum_{a\neq b}\left[Z(2\epsilon_{ab}^u+\epsilon_{ab}^d) 
    + (A-Z)(\epsilon_{ab}^u+2\epsilon_{ab}^d)\right]^2\ .
\end{align}
The new parameter dependence can lead to flavor-dependent cross sections. An incoming flavor state $\nu_a$
can produce either the same flavor state or an orthogonal one $\nu_b$. The first term in
Eq. (\ref{eq:NSI-change-xsec}) accounts for $\nu_a\to \nu_a$ scattering, while the second to scattering
to a flavor orthogonal state. Using the LBNF beamline, three NSI couplings---per first generation
quarks---can therefore be tested: $\epsilon_{\mu\mu}^q$, $\epsilon_{e\mu}^q$ and $\epsilon_{\mu\tau}^q$.

\begin{figure*}
    \centering
    \includegraphics[scale=0.336]{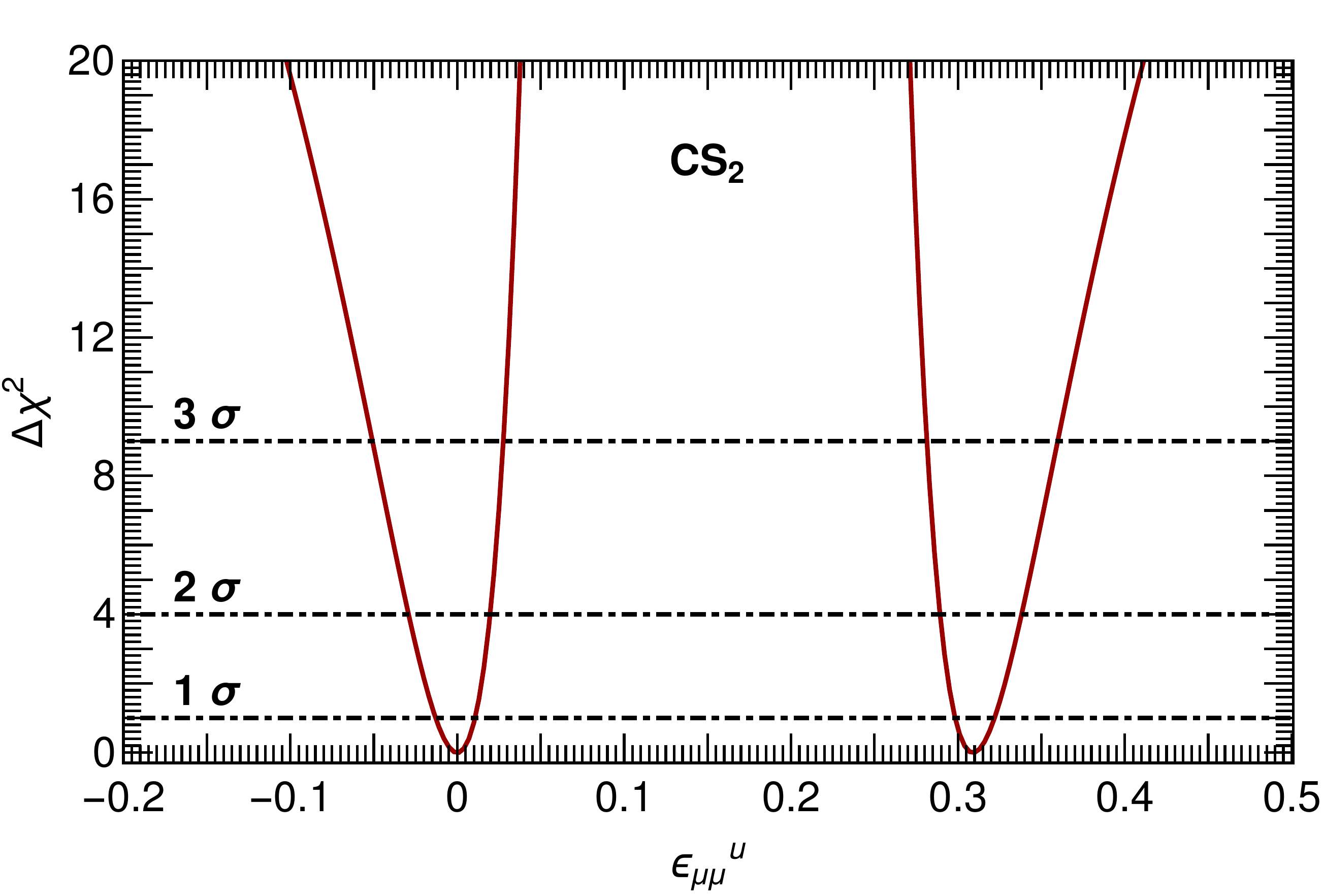}
    \includegraphics[scale=0.33]{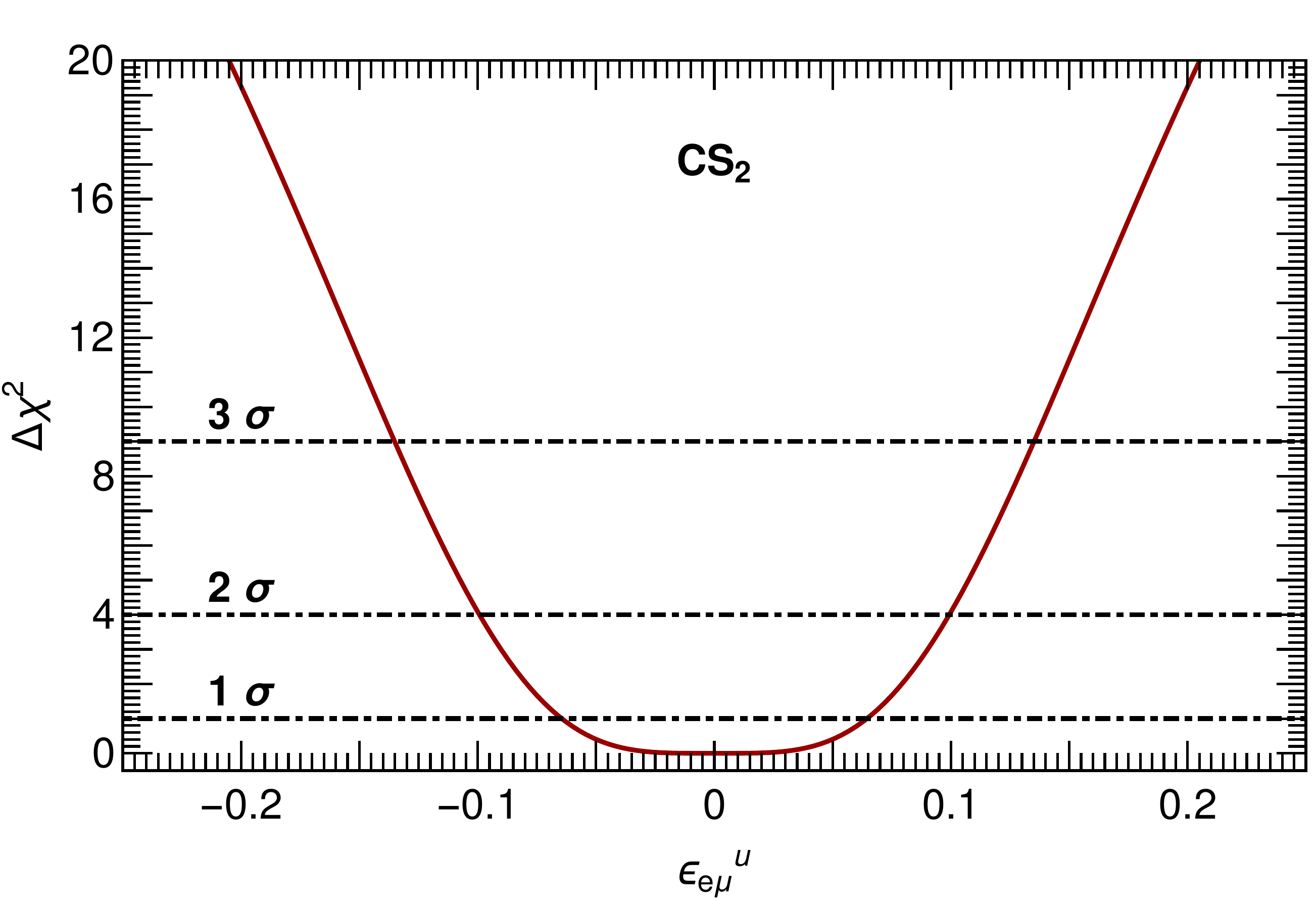}
    \caption{\textbf{Left graph}: Chi-Square distribution for $\epsilon_{\mu\mu}^u$ assuming
    the background-free hypothesis. Deviations due to background (10\% and 25\% of the signal rate) 
    are small. For the calculation we have assumed 100\% of the
    ten-cubic meter detector volume is filled with CS$_2$ and 7-years of data taking.
    \textbf{Right graph}: Same as left graph but for the off-diagonal coupling $\epsilon_{e\mu}^u$ (or $\epsilon_{\mu\tau}^u$). Results for down quark parameters are rather close to those
    found in this case and so are not displayed.}
    \label{fig:nsi_chiSq}
\end{figure*}

Calculation of sensitivities is done assuming one parameter at a time.
A procedure that is justified by the fact that is for this parameter
configurations for which the best sensitivies can be derived. In all 
cases we vary the effective parameter in the interval $[-1.0,1.0]$.
The results of the analysis are shown in Fig. \ref{fig:nsi_chiSq}.
Left graph for $\epsilon_{\mu\mu}^u$ and right graph for $\epsilon_{e\mu}^u$ 
(results for down quark couplings follow closely those for up quark 
parameters, so are not shown). Note that due to the adopted 
single-parameter analysis results for  $\epsilon_{\mu\tau}^u$ are identical 
to those from $\epsilon_{e\mu}^u$. Table \ref{tab:NSI-summary} summarizes the $1\sigma$ sensitivities that can be achieved along with $1\sigma$ intervals derived using COHERENT CsI spectral and timing information \cite{Giunti:2019xpr}.

For the flavor-diagonal coupling we find two disconnected allowed regions,
a result which is expected. The region around zero---which includes the
SM solution $\epsilon_{\mu\mu}^u=0$---is open just because contributions from the NSI
parameter generate small deviations from the SM prediction. The region of
large NSI---which does not include the SM solution---is viable because the NSI and
SM contributions destructively interfere, with the NSI contribution exceeding in about
a factor 2 the SM terms resulting in $-|Q_W|\to Q_W$. For the off-diagonal coupling
results are as well as expected. Since it contributes constructively enhancing the
SM prediction, the chi-square distribution is symmetric around $\epsilon_{e\mu}^u=0$.
Compared with results derived using COHERENT CsI spectral and timing information,
one can see that in all cases sensitivities improve. For $\epsilon_{\mu\mu}^u$
sensitivities are better by about a factor 3 (left interval) and 1.3 (right interval).
For $\epsilon_{\mu\mu}^u$ they improve by about a factor 2. These numbers apply as
well to the other NSI parameters not displayed. All in all one can see that
$\nu$BDX-DRIFT data will allow test region of NSI parameters not yet covered by
COHERENT measurements. 
\section{Conclusions}
\label{sec:conclusions}%
In this paper a new idea to study CE$\nu$NS with the $\nu$BDX-DRIFT detector has been considered. We have quantified sensitivities to the weak mixing angle using carbon disulfide as target material. Our findings demonstrate that a determination of this parameter at a renormalization scale within $\sim[0.1,0.4]\,$GeV can be done at the $8\%$ level, thus providing complementary information to future measurements at DUNE using the electron recoil channel. We have investigated as well sensitivities to the neutron distributions of carbon, fluorine and lead using carbon tetrafluorine and tetraethyllead as target materials. Our results show that measurements with accuracies of order $5\%$ and $10\%$, respectively, can be achieved. Finally, we have assessed sensitivities to new physics searches and for that aim we have considered effective neutrino NSI. Given the incoming neutrino flavor, at $\nu$BDX-DRIFT only muon-flavor NSI parameters can be tested. Using carbon disulfide as target material, flavor-diagonal (off-diagonal) couplings of order $10^{-3}$ ($10^{-2}$) can be proven. In the absence of a signal these numbers will translate in significant improvements of current limits.

Due to its directional and background rejection capabilities, the $\nu$BDX-DRIFT detector combined with the LBNF beamline provides a unique opportunity to study CE$\nu$NS in a neutrino energy range not yet explored. We estimated  the  ratio of the most important beam related neutrino-induced neutron background to the CE$\nu$NS signal to be small, about a factor  23 smaller. The detector offers a rich neutrino physics program---along with a potential agenda for light DM searches---that includes measurements of the CE$\nu$NS cross section in nuclides not used by other techonologies, measurements of the weak mixing angle in an energy regime not yet explored by any other neutrino scattering experiment, measurements of neutron distributions as well as searches for new physics in the neutrino sector.

\section*{Acknowledgments}
We thank Dinesh Loomba for discussions since the early stages of this work as well as for his suggestions on the manuscript. We thank Phil Barbeau, Pedro Machado and Kate Scholberg for comments on the manuscript. DAS is supported by the grant ``Unraveling new physics in the high-intensity and high-energy frontiers'', Fondecyt No 1171136. BD and LES acknowledge support from DOE Grant de-sc0010813. The work of DK is supported by DOE under Grant No. DE-FG02-13ER41976/ DE-SC0009913/DE-SC0010813.
\bibliography{references}
\end{document}